\documentclass[fleqn,usenatbib]{mnras}

\usepackage[T1]{fontenc}

\DeclareRobustCommand{\VAN}[3]{#2}
\let\VANthebibliography\thebibliography
\def\thebibliography{\DeclareRobustCommand{\VAN}[3]{##3}\VANthebibliography}

\usepackage{graphicx}	
\usepackage{amsmath}	
\usepackage{amssymb}	

\usepackage{newtxtext,newtxmath}

\title[Spiral structure of UGC\,4599]{The haloes and environments of nearby galaxies (HERON) -- III. A 45 kpc spiral structure in the GLSB galaxy UGC\,4599}
\author[A. Mosenkov et al.]{
Aleksandr V. Mosenkov,$^{1,2}$\thanks{E-mail: aleksandr\_mosenkov@byu.edu}
R. Michael Rich,$^{3}$
Michael Fusco,$^{4}$
Julia Kennefick,$^{4,5}$
David Thilker,$^{6}$
\newauthor
Alexander Marchuk,$^{2,13}$
Noah Brosch,$^{7}$
Michael West,$^{8}$
Michael Gregg,$^{9}$
Francis Longstaff,$^{10}$
\newauthor
Andreas J. Koch-Hansen,$^{11}$
Shameer Abdeen, $^{12}$
and William Roque$^{1}$
\\
\\
$^{1}$Department of Physics and Astronomy, N283 ESC, Brigham Young University, Provo, UT 84602, USA\\
$^{2}$Central (Pulkovo) Astronomical Observatory, Russian Academy of Sciences, Pulkovskoye Chaussee 65/1, St Petersburg 196140, Russia\\
$^{3}$Dept. of Physics and Astronomy, University of California, Los Angeles, CA 90095, USA\\
$^{4}$Arkansas Center for Space and Planetary Sciences, University of Arkansas, Fayetteville, AR 72701, USA\\
$^{5}$Department of Physics, University of Arkansas, Fayetteville, AR 72701, USA\\
$^{6}$Department of Physics and Astronomy, Johns Hopkins University, 3400 N. Charles Street, Baltimore, MD 21218, USA\\
$^{7}$The Wise Observatory and the Raymond and Beverly Sackler School of Physics and Astronomy, The Faculty of Exact Sciences,
Tel Aviv University, Tel Aviv 69978, Israel\\
$^{8}$Lowell Observatory, Flagstaff, AZ 86001, USA\\
$^{9}$Department of Physics, University of California, Davis, CA 95616, USA\\
$^{10}$University of California, Los Angeles, CA 90024, USA\\
$^{11}$Zentrum f{\"u}r Astronomie der Universit{\"a}t Heidelberg, Astronomisches Rechen-Institut, M\"onchhofstr. 12, Heidelberg 69120, Germany\\
$^{12}$Dept. of Physics and Astronomy, Georgia Southern University, GA 30458, USA\\
$^{13}$Saint Petersburg State University, Universitetskij pr. 28, St. Petersburg 198504, Russia\\
}

\date{Accepted XXX. Received YYY; in original form ZZZ}

\pubyear{2022}

\begin{document}
\label{firstpage}
\pagerange{\pageref{firstpage}--\pageref{lastpage}}
\maketitle

\begin{abstract}

We use a 0.7-m telescope in the framework of the Halos and Environments of Nearby Galaxies ({\sl HERON}) survey to probe low surface brightness structures in nearby galaxies. One of our targets, UGC\,4599, is usually classified as an early-type galaxy surrounded by a blue ring making it a potential Hoag's Object analog. Prior photometric studies of UGC\,4599 were focused on its bright core and the blue ring. However, the {\sl HERON} survey allows us to study its faint extended regions. With an eight hour integration, we detect an extremely faint outer disk with an extrapolated central surface brightness of $\mu_\mathrm{0,d}(r)=25.5$~mag\,arcsec$^{-2}$ down to 31~mag\,arcsec$^{-2}$ and a scale length of 15~kpc. We identify two distinct spiral arms of pitch angle $\sim6\degr$ surrounding the ring. The spiral arms are detected out to $\sim45$~kpc in radius and the faint disk continues to $\sim70$~kpc. These features are also seen in the GALEX FUV and NUV bands, in a deep $u$-band image from the 4.3m Lowell Discovery Telescope (which reveals inner spiral structure emerging from the core), and in H{\sc i}. We compare this galaxy to ordinary spiral and elliptical galaxies, giant low surface brightness (GLSB) galaxies, and Hoag's Object itself using several standard galaxy scaling relations. We conclude that the pseudobulge and disk properties of UGC\,4599 significantly differ from those of Hoag's Object and of normal galaxies, pointing toward a GLSB galaxy nature and filamentary accretion of gas to generate its outer disk.
\end{abstract}




\section{Introduction}

UGC\,4599 is a nearby galaxy (often classified as S0) with an unusual ring morphology similar to that observed in Hoag's Object (see Table~\ref{tab:generals} for the general properties of this galaxy). \citet{2011MNRAS.413.2621F} identify it as the nearest target of its kind, and, therefore, a promising candidate to explore the class of Hoag-type galaxies. Hoag's Object \citep{1950AJ.....55Q.170H} is characterized by a red central spherical body (an elliptical galaxy or a bulge) surrounded by a distinctive blue star-forming ring. Hoag's Object was also found to contain an H{\sc i} ring 1.5 times more extended than the star-forming ring \citep{2013MNRAS.435..475B}. 

Formation of ring galaxies may go through a variety of possible pathways. These include secular evolution through bar-related resonances \citep{1996FCPh...17...95B}, gas-rich accretion from other galaxies \citep[see e.g.,][and references therein]{1983AJ.....88..909S,2020A&A...638L..10S}, gas accretion from the intergalactic medium via cosmological filaments \citep{2006ApJ...636L..25M}, wet minor mergers \citep{1996FCPh...17...95B, 1997A&A...325..933R}, and collisions of two or more galaxies \citep{1976ApJ...209..382L, 1996FCPh...16..111A}. In particular, polar-ring galaxies can be formed by orthogonal or at least significantly non-coplanar galactic collisions \citep{1983AJ.....88..909S,2003A&A...401..817B}. Roughly 0.5\% of nearby S0 galaxies are shown to contain polar rings (\citealt{1990AJ....100.1489W}, \citealt{2011MNRAS.418..244M}). 

In their detailed study of UGC\,4599, \citet{2011MNRAS.413.2621F} decomposed UGC\,4599 into a reddish de Vaucouleurs core, a blue ring (with some evidence of spiral structure), and detected an extended H{\sc i} disk. The imaging of both \citet{2011MNRAS.413.2621F} and \citet{2011AJ....142..145G} reveals signatures of an outer low-surface brightness (LSB), star forming disk. Therefore, we decided to select this target for deep LSB imaging with the 0.7m Jeanne Rich telescope as one of the earliest targets of the Haloes and Environments of Nearby galaxies ({\sl HERON}) survey \citep[see][]{2019MNRAS.490.1539R}. 

As ground-based telescopes develop in observing capabilities, both through the use of single small-aperture telescopes \citep[see e.g.,][]{2010AJ....140..962M,2019MNRAS.490.1539R} and arrays \citep{2014PASP..126...55A,2019arXiv191111579S}, the identification and characterization of LSB objects has become more common. For example, the use of very deep photometric observations resulted in the discovery of many exceedingly diffuse and ultra diffuse galaxies \citep{2015ApJ...798L..45V,2015ApJ...807L...2K,2016ApJ...828L...6V,2018Natur.555..629V}.  

Giant LSB (GLSB) galaxies also require long exposures to study their outskirts. They host very faint and large star-forming disks up to 250~kpc in diameter \citep[see e.g.,][]{2016A&A...593A.126B} which are characterized by massive dark matter halos \citep{2013JApA...34...19D}, high hydrogen gas surface densities \citep{1987AJ.....94...23B}, and generally low metallicities \citep{2010MNRAS.409..213L}. Their extended disks have extrapolated central surface brightness values much lower than the \citet{1970ApJ...160..811F} value of $\mu_\mathrm{0,d}(B)= 21.65$~mag\,arcsec$^{-2}$ for disk galaxies. GLSB galaxies usually have a high surface brightness bulge \citep{2021MNRAS.503..830S}, loosely wound spiral arms \citep{1997PASP..109..745B}, and generally lack bars. However, in GLSB galaxies with a bulge-dominated central component, tighter spiral arms may be present \citep{2013JApA...34...19D}. GLSB galaxies also sometimes demonstrate ring structures, as in the case of UGC\,6614 \citep{2013JApA...34...19D}. 

The formation histories of GLSB galaxies are not well understood but include catastrophic collisional scenarios, involving (1) merging \citep{2018MNRAS.481.3534S,2023MNRAS.523.3991Z}, (2) head-on collisions with a massive galaxy leading to the collisional formation of a ring galaxy \citep{2008MNRAS.383.1223M}, (3) the accretion of gas-rich low-mass galaxies \citep{2006ApJ...650L..33P,2016ApJ...826..210H}, or (4) the accretion of the cooling hot halo gas stimulated by the merger of intruding galaxies \citep{2018MNRAS.480L..18Z}. Also, non-catastrophic solutions have been proposed: (1) isolated secularly evolved models \citep{2001MNRAS.328..353N,2016A&A...593A.126B}, (2) the presence of an unusual shallow and extended dark matter halo \citep{2014MNRAS.437.3072K} which can lead to the formation of a giant LSB disk; (3) or the cold accretion of gas from the intergalactic medium \citep{2010A&A...516A..11L,2019MNRAS.489.4669S}. These scenarios have been recently investigated by \citet{2021MNRAS.503..830S} for a sample of seven well-known GLSB galaxies. They conclude that the GLSB galaxies represent a heterogeneous class of giant galaxies with LSB disks. This suggests that their formation can involve various aforementioned mechanisms, among which the external supply of material to form the giant LSB disk may be more common.

The aim of this paper is to explore the structural properties of UGC\,4599 in great detail using deep imaging and compare its structure with ordinary spiral and elliptical galaxies, Hoag's  Object, and with GLSB galaxies employing standard galaxy scaling relations. This comparison aims to propose a plausible scenario for the formation of the peculiar structure of UGC\,4599. We confirm recent results by \citet{2023A&A...669L..10S} and identify UGC\,4599 as a GLSB galaxy with a high surface brightness central component surrounded by a blue star-forming ring and hosting a LSB disk with barely detectable embedded spiral structure.
Using supplementary observations in the $u$ waveband from the 4.3m Lowell Discovery Telescope, we also reveal a dim spiral structure within the galaxy core which excludes the possibility that the central body of UGC\,4599 is an elliptical galaxy, unlike Hoag's Object.  

We organize the paper as follows. Sections~\ref{sec:observations} and~\ref{sec:processing} describe our observations and image processing. Section~\ref{sec:analysis} details our photometric analysis of the galaxy at different wavelengths including investigation of its spiral structure. In Section~\ref{sec:discussion}, we discuss the structural properties of UGC\,4599 compared to other galaxies using famous galaxy scaling relations and propose realistic formation scenarios of this galaxy. Section~\ref{sec:conclusions} summarizes our conclusions. Throughout the paper, all magnitudes and surface brightnesses are given in the AB magnitude system.

\begin{table}
 \centering
 \begin{minipage}{75mm}
  \centering
  \parbox[t]{75mm} {\caption{General properties of UGC\,4599.}
  \label{tab:generals}}
  \begin{tabular}{lcc}
  \hline 
  \hline
RA (J2000) & 08$^\mathrm{h}$47$^\mathrm{m}$41$^\mathrm{s}$ & (1)  \\ 
Dec. (J2000) & +13$^\mathrm{d}$25$^\mathrm{m}$09$^\mathrm{s}$ & (1)  \\
Stellar heliocentric velocity (km\,s$^{-1}$) & 2071 & (2) \\
Distance (Mpc) & 32.0 & (3) \\
Scale (kpc\,arcsec$^{-1}$) & 0.156 & (3)  \\
Type & (R)SA0 & (1) \\
Major diameter (arcmin) & 2.1 & (1) \\
$m_r$ (mag) & 13.47 & (4) \\
$M_r$ (mag) & -19.06 & (4) \\
$g-r$ & 0.67 & (4) \\
  \hline\\
  \end{tabular}
  \end{minipage}
     
     \parbox[t]{85mm}{ {\bf References:}
     (1) NASA/IPAC Extragalactic Data base (NED);
     (2) \citet{2020AJ....160..271D};
     (3) \citet{2018ApJ...861...49H};
     (4) {\it PhotoObj} Sloan Digital Sky Survey (SDSS, \citealt{2020ApJS..249....3A}) table. Apparent and absolute magnitudes in the $r$ band and the $g-r$ color were corrected for Galactic extinction using \citet{2011ApJ...737..103S} and K-correction using the K-corrections calculator \citep{2010MNRAS.405.1409C,2012MNRAS.419.1727C}.}
 \end{table} 

\section{Observations}
\label{sec:observations}	

\begin{figure*}
	\includegraphics[width=\textwidth]{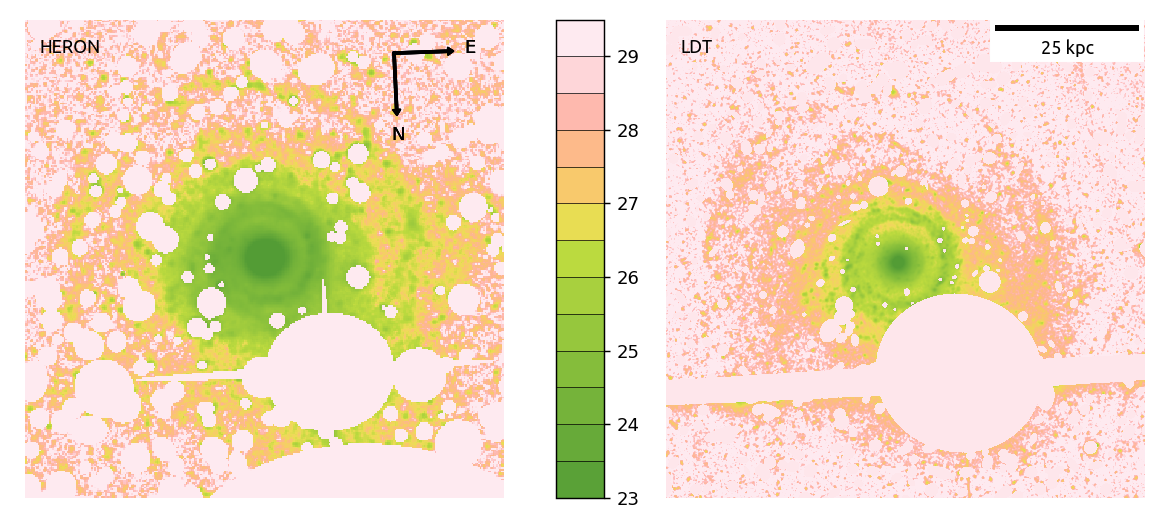}
	\includegraphics[width=\textwidth]{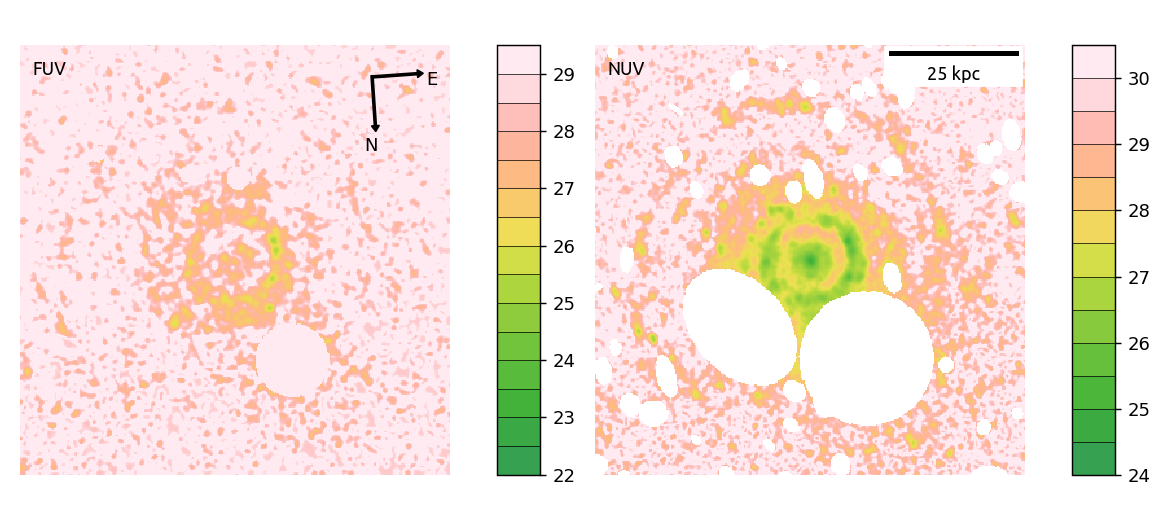} 
	\caption{Images of UGC\,4599 in the {\sl HERON} Luminance filter converted to the SDSS $r$ band ({\it top left}) and in the LDT $u$ band ({\it top right}). The bottom panels show the supplementary GALEX FUV ({\it bottom left}) and GALEX NUV ({\it bottom right}) images. The colorbars decode the surface brightnesses in magnitudes per square arcsec depicted in the different images. Stars and other background and foreground objects are masked out. The masked west ellipse in the NUV band covers an image artifact.}
	\label{surf_map}
\end{figure*}

Deep optical observations of UGC\,4599 were obtained in January 2011, with 0.7-m Jeanne Rich Telescope Centurion 28 located near Frazier Park, CA, northwest of Los Angeles. 
The detector at f/3.2 prime focus provides a 2-3~arcsec resolution. 
On a moonless night, the zenith sky brightness at the site is typically $\sim 21.6-22.1$~mag\,arcsec$^{-2}$. A sister telescope of identical design was implemented later at the Wise Observatory in Israel and is described in \citet{2015Ap&SS.359...49B}. 
	
In the {\sl HERON} survey, we make use of the Luminance (L) band filter that allows us to reduce the night sky background and artificial light pollution. Using this Luminance filter, we are able to obtain very deep images of local galaxies and explore their LSB structure with the aid of a relatively small-aperture telescope. 
	
By employing the wide $4000-7000$\AA\, Luminance filter for UGC\,4599 with an 8 hour integration, we rich a surface brightness limit of 30.2~mag\,arcsec$^{-2}$ calibrated to the Sloan Digital Sky Survey (SDSS) $r$ filter (as a $3\sigma$ level in square boxes of $10\arcsec \times 10\arcsec$).
The transformation to SDSS has a negligible color term and is straightforward using calibration stars in the field \citep[see][]{2019MNRAS.490.1539R}. For extended surface brightness regions, it is possible to detect structures marginally fainter than the 31~mag\,arcsec$^{-2}$ limit by averaging intensities in grouped pixels (see Sect.~\ref{sec:profiles}). The average point spread function (PSF) FWHM for our stacked image is 1.9~arcsec at a pixel size of 1.67~arcsec.
	
Supplementary observations of UGC\,4599 were conducted on the night of November 29, 2016, with the 4.3-m Lowell Discovery Telescope (LDT), formerly the Discovery Channel Telescope (DCT). This observatory is located forty miles southeast of Flagstaff, AZ. To acquire images, we used the Large Monolithic Imager (LMI) with a single $6144 \times 6160$ e2v CCD. All images were binned $2 \times 2$, so that the pixel scale is 0.24~arcsec and the field of view is $12.3 \times 12.3$~arcmin. We took five individual exposures, each of 1200s duration, using the SDSS $u$ filter. To remove the effects of bad pixels and artifacts, as well as to improve flat fielding, the exposures were obtained with dithering. Average seeing was poor (FWHM=1.7~arcsec) due to some thin cirrus contamination. To obtain co-add images, we used the same standard data reduction techniques as for the {\sl HERON} data. The depth of the final stacked image occurred to be 30.1~$\mathrm{mag\,arcsec}^{-2}$ ($3\sigma$ in $10\arcsec \times 10\arcsec$ boxes, the $u$ band).

We also use supplementary observations from the GALEX satellite in the FUV band centered at $1450$\AA\, (the AIS survey with a total exposure of 217s) and NUV band centered at $2300$\AA\, (the MIS survey with a total exposure of 2720s) using the last publicly available data release\footnote{\url{http://galex.stsci.edu/gr6}}. The advantage of using UV observations is that they have less contamination from foreground stars than in the optical. We also make use of the Infrared Array Camera (IRAC, \citealt{2004ApJS..154...10F}) observations at 3.6~$\mu$m (331s of duration) from the {\it Spitzer} Space Telescope \citep{2004ApJS..154....1W} using the {\it Spitzer} Heritage Archive\footnote{\url{https://sha.ipac.caltech.edu/}}. Finally, we use VLA observations from \citet{2010PhDT.......102D} for mapping the H{\sc i} distribution in UGC\,4599.

\section{Image processing}
\label{sec:processing}		

Initial image reduction was carried out for both the {\sl HERON} and LDT observations of UGC\,4599 using the standard IRAF routines (bias subtraction, correction for the dark current, flat fielding, etc.). 

For each image, we estimated the sky background around the galaxy within an elliptical annulus of radius 500~arcsec (this radius was chosen based on the extent of the galaxy after a preliminary sky subtraction for all the images) and a width of 30~arcsec. 

Due to the presence of several bright stars in our deep exposures, some further image preparation was done which is of high importance for an analysis of the faint outer structure of UGC\,4599. Instead of simple masking, we used the IRAF routine {\tt{DAOPHOT}} to subtract the stars. Although careful star subtraction is a relatively time consuming process, it is worthwhile here to retain more of the structure of the galaxy than while using simple masking techniques. In our modelling, we take into account the core of the PSF determined for non-saturated bright stars found in the field near the galaxy and the wings of the extended PSF built in \citet{2019MNRAS.490.1539R} (see their fig.~10) for the {\sl HERON} survey. For the LDT frame, we used the brightest saturated star BD+13\,1990 near the galaxy ring. 
However, even a relatively good star subtraction with {\tt{DAOPHOT}} did not allow us to completely remove the light of the foreground stars, especially those that are saturated: such stars exhibit diffraction spikes and saturated cores, so a good PSF subtraction is not possible in this case. Unfortunately, the bright foreground star BD+13\,1990 completely obscures an interesting region of the galaxy that likely contains a large segment of the spiral structure under study. The GALEX and LDT UV images are useful for studying this segment, since the aforementioned star is not as bright in the UV as in the optical. 

To completely eliminate all contaminating sources in our 1D fitting and 2D photometric decompositions (see Sects.~\ref{sec:profiles} and \ref{sec:decomposition}) and to examine the spiral structure (Sect.~\ref{sec:spiral_structure}), we created a mask for each of the images by employing {\tt{SExtractor}} \citep{1996A&AS..117..393B} and then manually revised it to mask off objects that do not belong to the target galaxy. The UV and optical images of UGC\,4599 with the masks superimposed are displayed in Fig.~\ref{surf_map}.

In order to describe the surface brightness distribution of the galaxy, an intensity map in a more common filter system than Luminance is required. For this, we calculated a conversion from the instrumental Luminance $L$ filter magnitude to the SDSS $r$ magnitude. We selected 18 non-saturated stars of known SDSS $g$ and $r$ magnitudes in the field of UGC\,4599. Aperture photometry of these stars was then performed using the IRAF routine {\tt{DAOPHOT}}. After that, the Luminance instrumental magnitudes of these stars were compared to the SDSS $g$ and $r$ magnitudes of the same objects for deriving a calibration equation to convert pixel counts in our image to values of $r$-mag\,arcsec$^{-2}$ for each pixel in the image. The $g-r$ term in the calibration equation appeared to be very small, especially if we take into account different kinds of errors. These errors originate from the photometric errors of the archived magnitudes, the errors in our instrumental photometry, and the fit errors in the comparison between our instrumental and the archival magnitudes.

\begin{figure}
\includegraphics[width=0.46\textwidth]{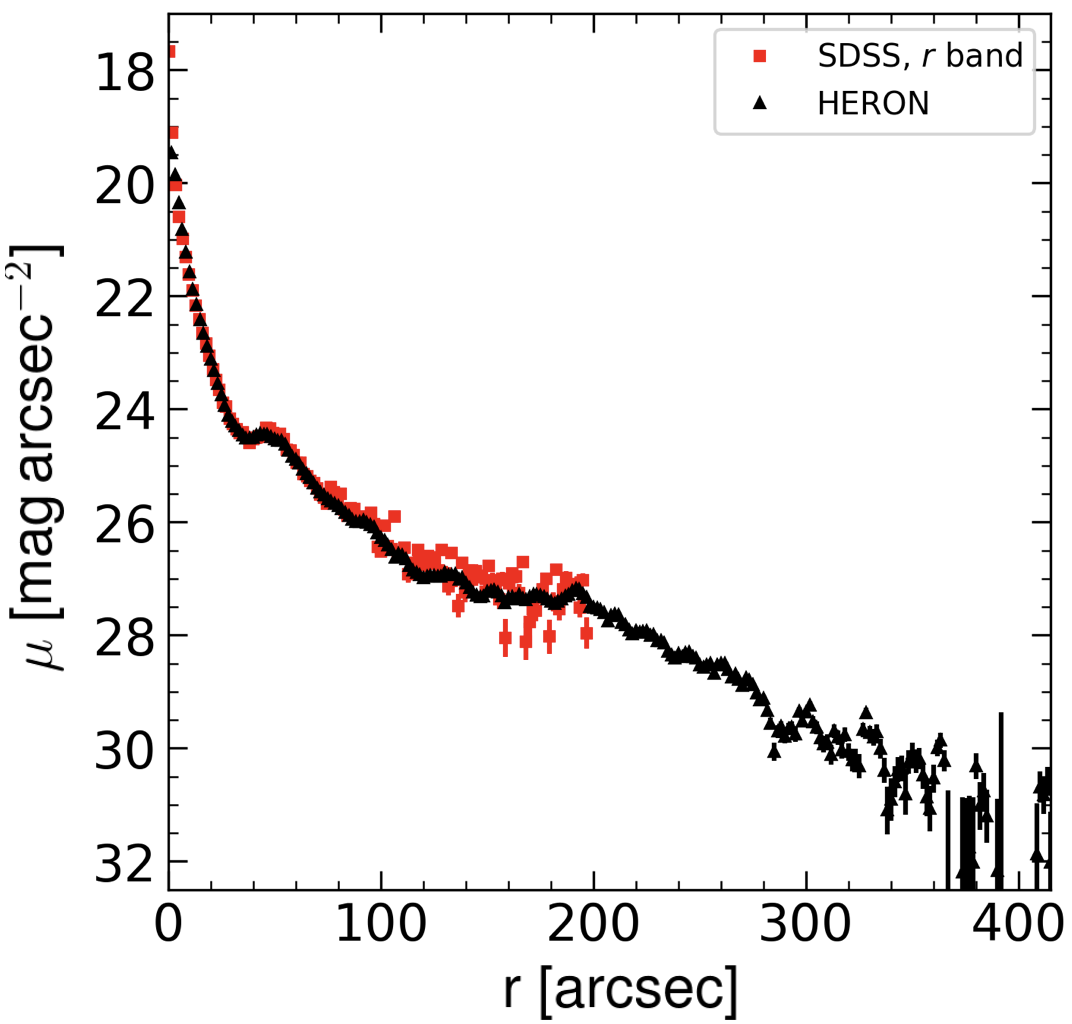}
\caption{Comparison between the azimuthally averaged profiles created for the {\sl HERON} $L$-band and SDSS $r$-band data.}
\label{compar_heron_sdss}
\centering
\end{figure}

In Fig.~\ref{compar_heron_sdss}, we show a comparison for the azimuthally averaged profiles created for the {\sl HERON} image with the above described $r$-band calibration and an $r$-band SDSS image of UGC\,4599. As one can see, the comparison is very good ($\Delta \mu<0.08$~mag\,arcsec$^{-2}$) within 100 arcseconds from the center and is gets worse ($\Delta \mu<0.26$~mag\,arcsec$^{-2}$) outwards where the depth of the SDSS photometry is reaching its limit ($26.5$~mag\,arcsec$^{-2}$, $3\sigma$ in $10\arcsec \times 10\arcsec$ boxes in the $r$ band).

\section{Structural analysis of UGC\,4599}
\label{sec:analysis}	

\subsection{Azimuthally averaged profiles}
\label{sec:profiles}

We use different data sources, described in Sect.~\ref{sec:observations}, to explore the surface brightness distribution in UGC\,4599 and trace its general structural components, the bulge and the disk. In particular, we exploit the GALEX FUV and NUV observations to study the distribution of the young (0-200~Myr) stellar population coupled with the deep LDT $u$-band image, which is sensitive to young stars with a larger range of ages. The deep {\sl HERON} coadded frame maps the distribution for a mix of the old and young stellar populations, whereas the {\it Spitzer} IRAC 3.6~$\mu$m image traces the old stellar population, which makes up the bulk of the galaxy stellar mass. Finally, we analyze the VLA H{\sc i} map to ascertain the distribution of the neutral hydrogen gas in the galaxy disk.

In this section, we carry out azimuthally averaged profile fitting of the aforementioned data sets, whereas a detailed 2D decomposition of the galaxy is presented in Sect.~\ref{sec:decomposition}. To facilitate a general galaxy structural analysis, 1D fitting is preferable to 2D fitting due to the low signal-to-noise ratio at the periphery of the galaxy where the outer disk dominates the galaxy profile.

To properly model the galaxy profiles, we take into account the PSF for each galaxy image used. For GALEX, the PSF kernels in the NUV and FUV wavebands were taken from \citet{2011PASP..123.1218A}. For the LDT $u$-band image, bright, non-saturated stars far from the galaxy were fitted with a Moffat function, and the PSF wings were approximated using the bright saturated star BD+13\,1990. An extended IRAC 3.6$\mu$m PSF was retrieved from the IRSA website\footnote{\url{https://irsa.ipac.caltech.edu/data/SPITZER/docs/irac/calibrationfiles/psfprf/}}. For VLA H{\sc i}, we used a synthesized beam size of 21.1~arcsec by 14.4~arcsec with a position angle of 8.2$\degr$. All PSF images were then azimuthally averaged to be taken into account in our 1D fitting of the galaxy profiles. 

\begin{figure*}
	\includegraphics[width=0.32\textwidth]{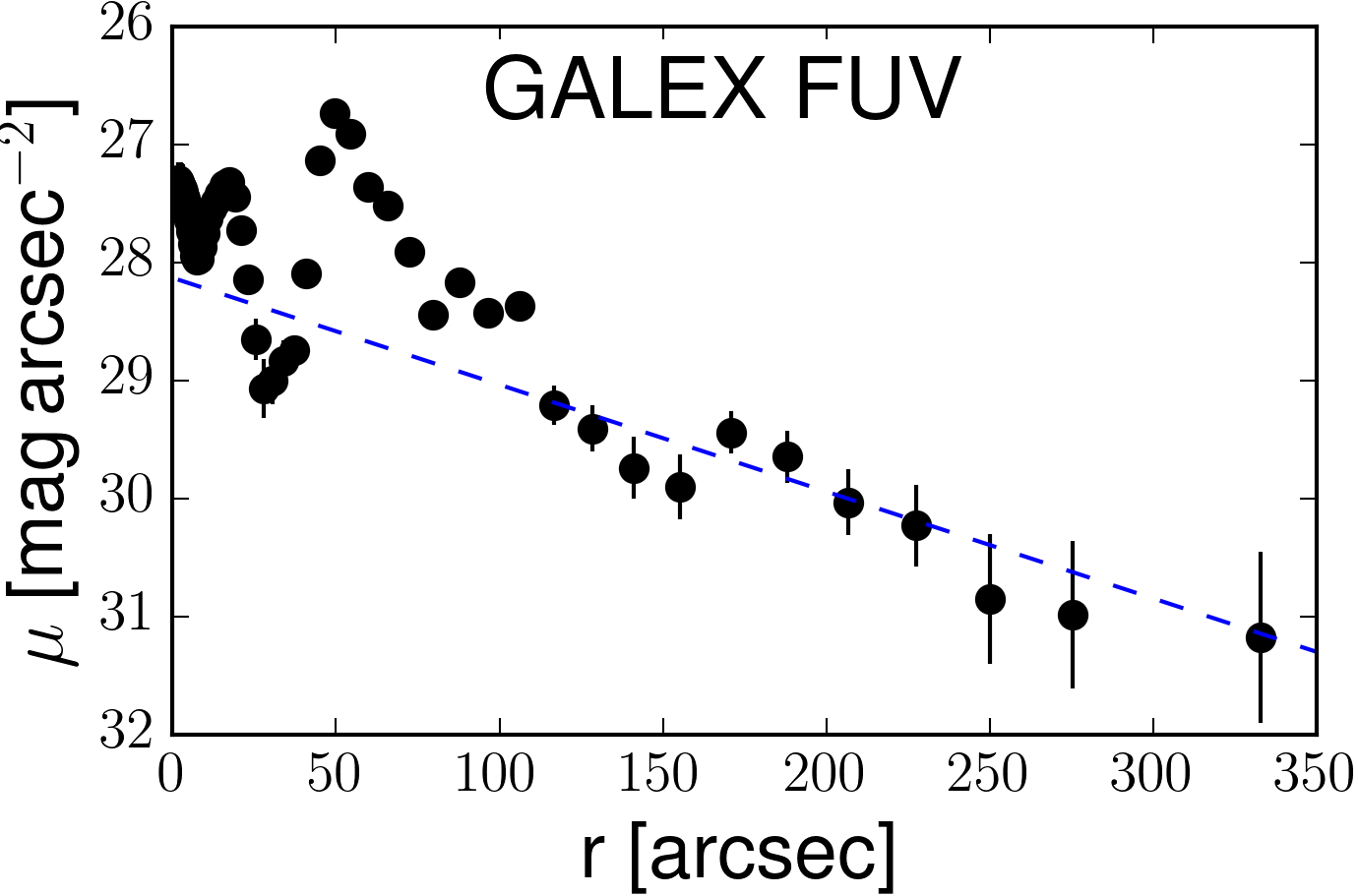}
	\includegraphics[width=0.32\textwidth]{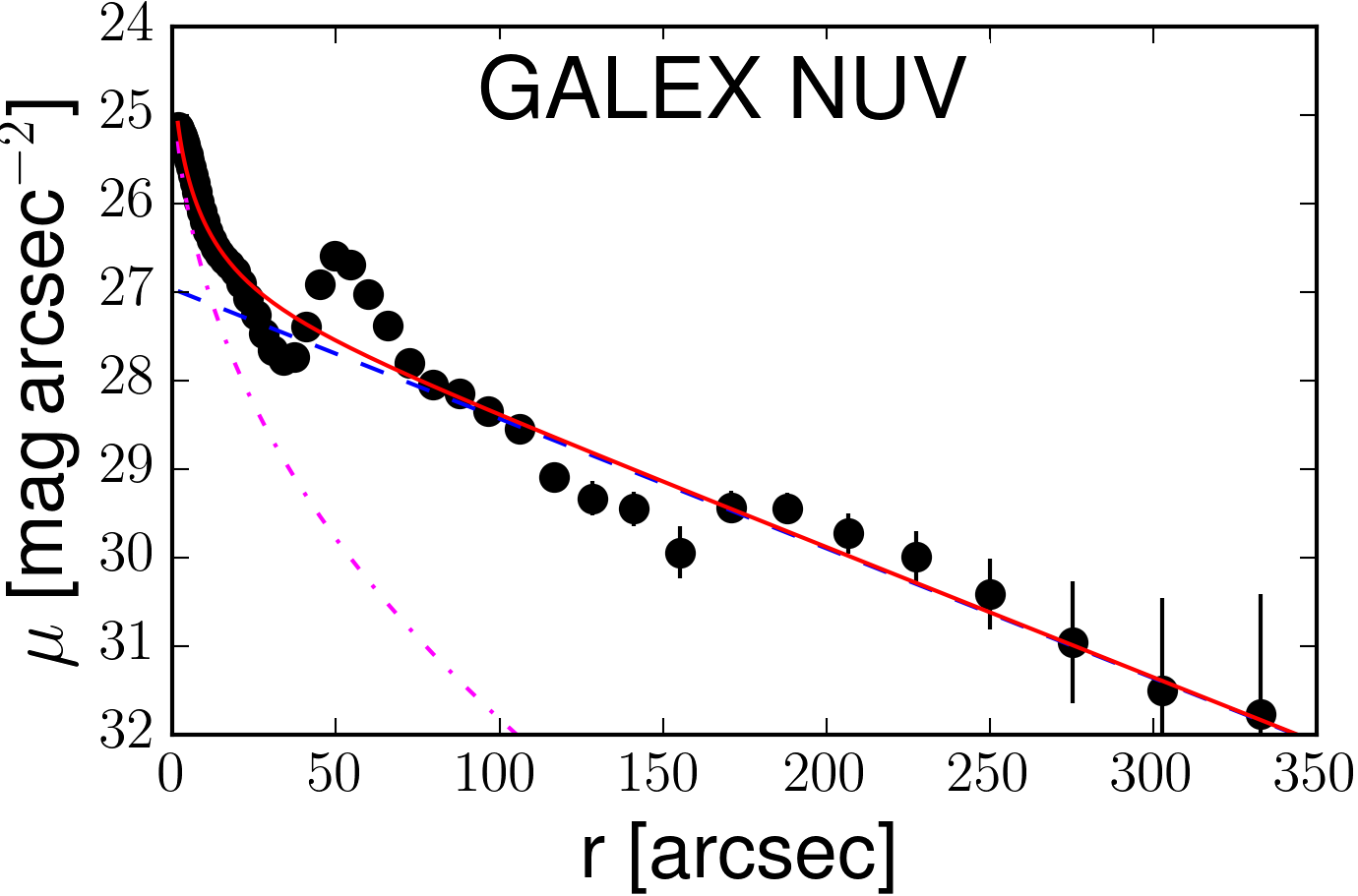}	
	\includegraphics[width=0.32\textwidth]{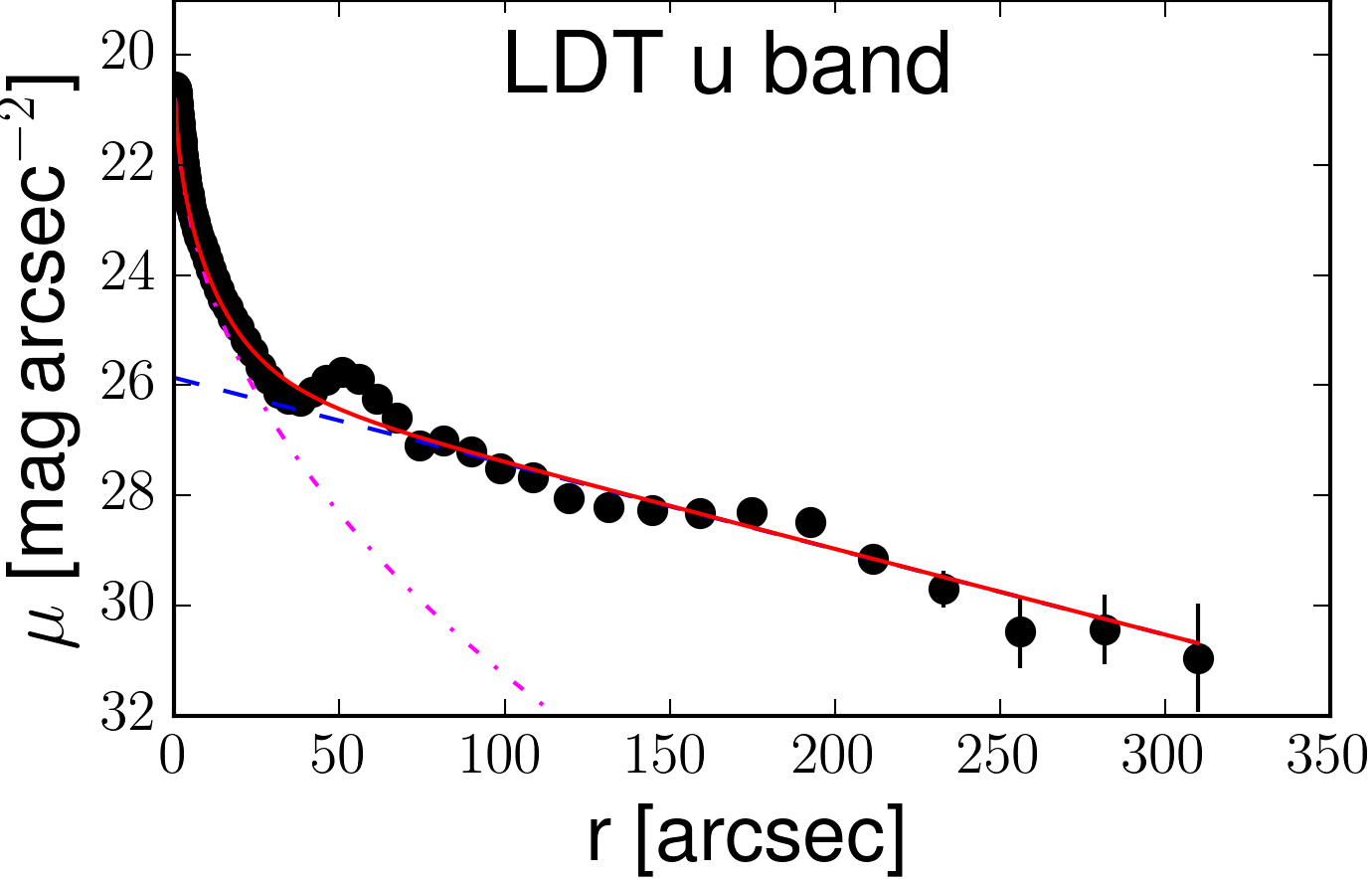}
	\includegraphics[width=0.32\textwidth]{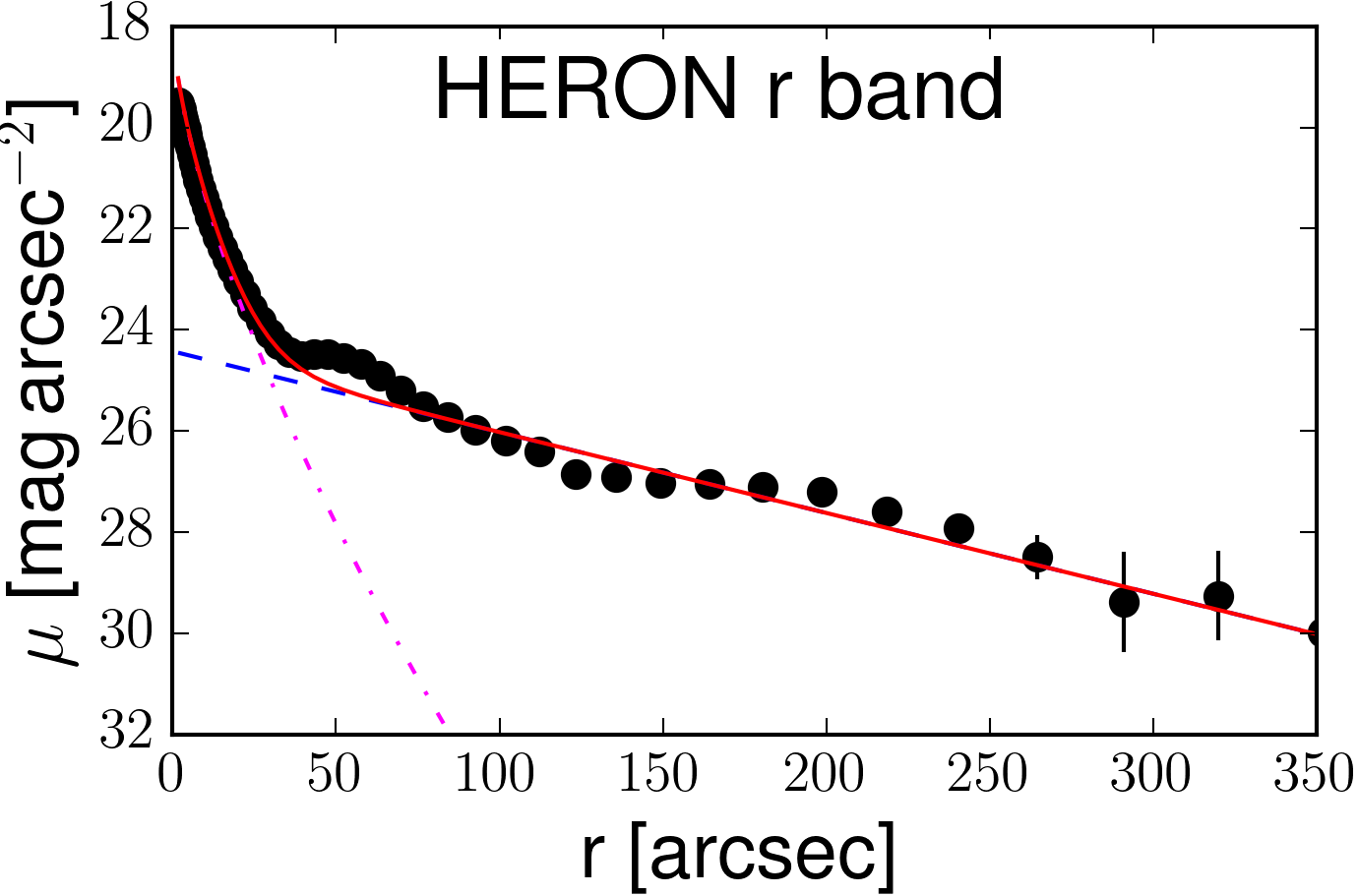}
	\includegraphics[width=0.32\textwidth]{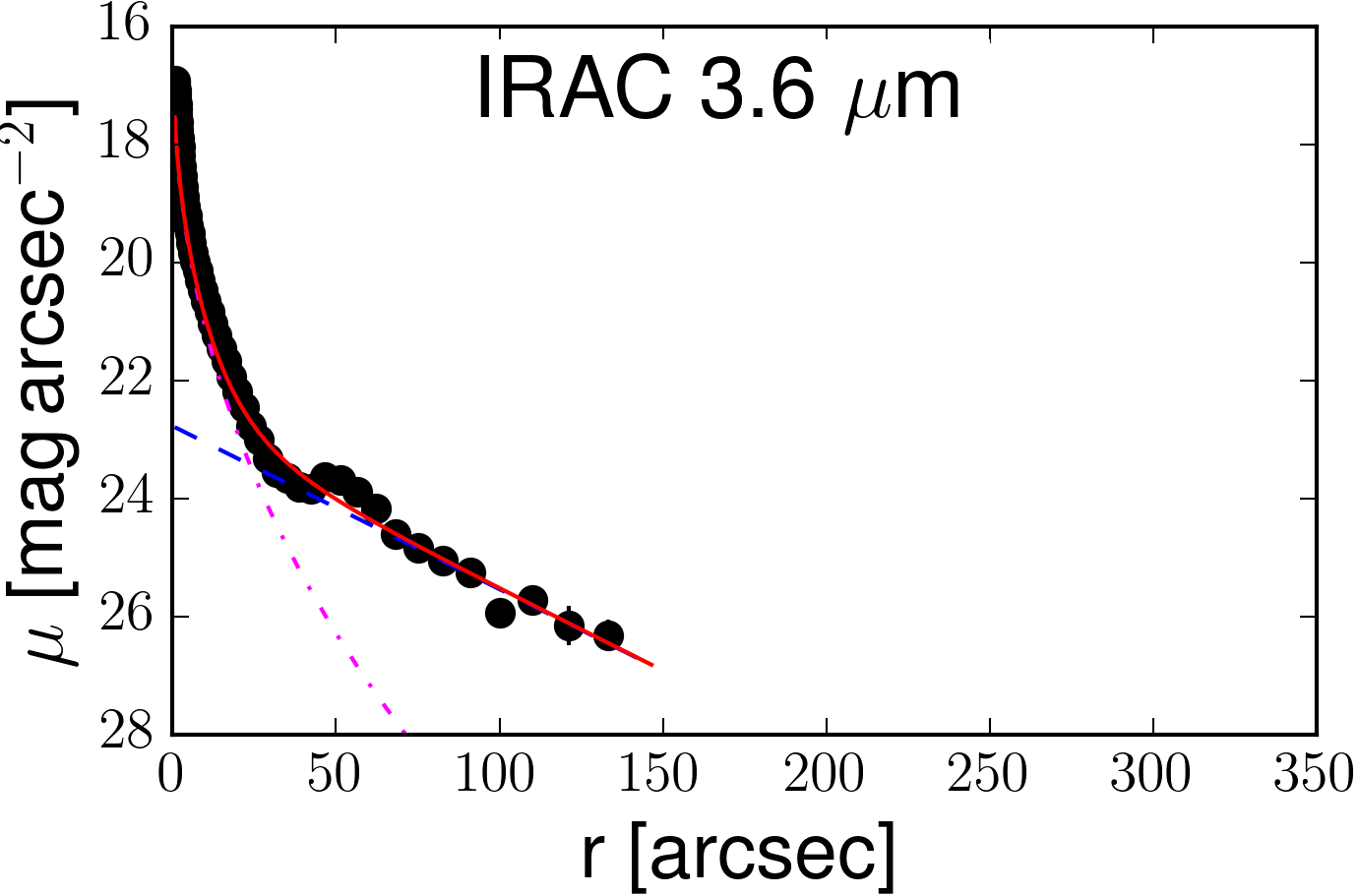}
	\includegraphics[width=0.32\textwidth]{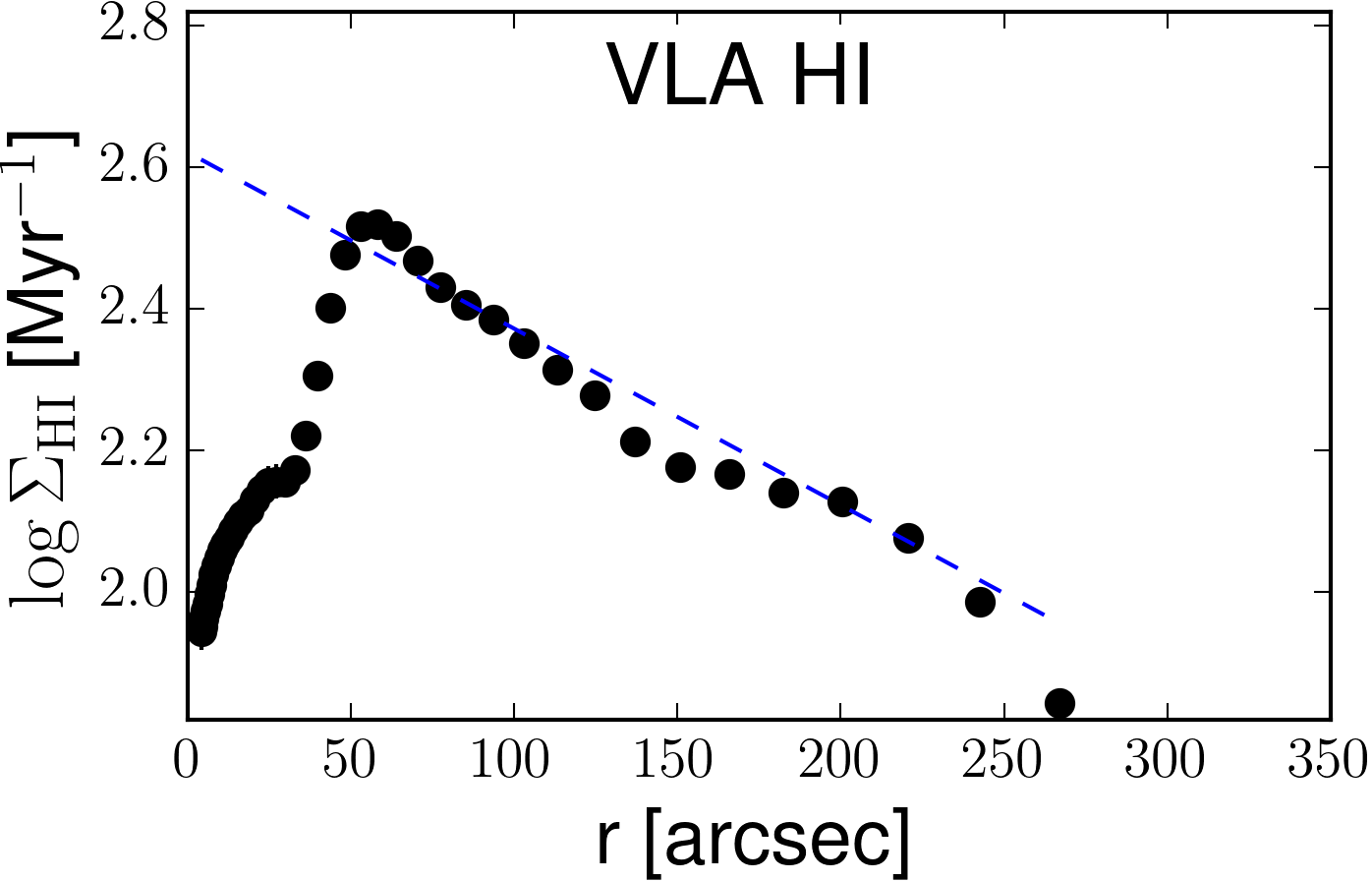}

	\caption{Profiles of UGC\,4599 with superimposed models: blue dashed lines correspond to the disk, magenta dashed lines -- to the bulge, and red lines depict the total model.}
	\label{azim_profiles}
\end{figure*}

In all the wavebands, except for FUV and 21~cm, a significant intensity peak is present coinciding with the center of the galaxy core. Therefore, in these bands, we use a `bulge+disk' model to accurately fit the overall 1D galaxy profile, whereas for the GALEX NUV and VLA H{\sc i} profiles we use a single-disk profile excluding an inner region within 75~arcsec. In all the cases, the disk profile was approximated using an exponential function \citep{1940BHarO.914....9P,1970ApJ...160..811F}, whereas the core (if bright enough) was fitted with a S\'ersic function \citep{1963BAAA....6...41S,1968adga.book.....S}. In all the profiles, the star-forming ring from $R_\mathrm{r,in}=30$~arcsec to $R_\mathrm{r,out}=75$~arcsec (where it dominates the surface brightness distribution) was masked off. Unfortunately, the IRAC 3.6$\mu$m field is too small to cover the entire galaxy, so only an inner part of the faint disk within $R\sim150$~arcsec is showcased. This, however, is sufficient to retrieve the parameters of the general distribution for the old stellar population in UGC\,4599. The created azimuthally averaged profiles, along with the superimposed models, are presented in Fig.~\ref{azim_profiles}. The parameters of the disk for each of the bands under consideration are listed in Table~\ref{tab:ProfilesFitting}.

We point out that although, to the first approximation, the surface brightness distribution beyond $r\approx75$~arcsec can be fitted with a single exponential profile relatively well, some deviations from this fit are apparent in Fig.~\ref{azim_profiles}, especially beyond 120~arcsec. Based on the scheme proposed in \citet{2005ApJ...626L..81E}, \citet{2006A&A...454..759P}, and \citet{2008AJ....135...20E}, \citet{2011AJ....142..145G} classify their $B$-band profile of UGC\,4599 as Type III-d: the outer profile is bending up (``antitruncation'') and it is part of the galaxy disk through the extended spiral structure. They determine a break of the disk profile at $r=121$~arcsec, beyond which the gradient of the surface brightness profile slightly changes and can be fitted with a different slope and intercept. However, as can be seen in Fig.~\ref{surf_map}, the surface brightness profile of the inner disk (dominating within $75<r<120$~arcsec) is significantly affected by the spiral arms wound out from the ring. As we show in Sect.~\ref{sec:spiral_structure}, the average pitch angle of the spiral structure is fairly small, so at smaller radii from the ring, the light from the spiral arms make a significant contribution to the galaxy luminosity profile and may cause it to appear ``higher'' over the outer disk profile extrapolated into the inner disk region. Therefore, a simple `bulge+disk' decomposition is beneficial because it allows one to determine the general parameters of the {\it main} galaxy components and compare them to those of other galaxies (see Sect.~\ref{sec:scaling_relations}) for which a similar simple decomposition has been performed, without going into detail of their structure. In addition, we note that the outer disk extends out to $\sim350$~arcsec and dominates across a five times larger range of radii than the inner disk. Obviously, it greatly affects the parameters of the general disk provided in Table~\ref{tab:ProfilesFitting}. At the same time, the profile of the inner disk with the smaller errorbars also non-negligibly contributes to the fitting, and, as a result, we measure an `average' galaxy disk.

\begin{table}
\caption{Best-fit parameters of the disk and bulge listed for different wavelengths. The central and effective surface brightnesses have been corrected for Galactic foreground extinction (using \citealt{2011ApJ...737..103S}) and K-correction \citep{2010MNRAS.405.1409C,2012MNRAS.419.1727C}. The S\'ersic index $n=2.4$ (found for the {\sl HERON} profile) is kept fixed for all models.}
\label{tab:ProfilesFitting}
\centering
    \begin{tabular}{lcccc}
    \hline
    \hline    
    Data &  $\mu_\mathrm{0,d}$ & $h$ & $\mu_\mathrm{e,b}$ & $r_\mathrm{e,b}$ \\[0.5ex]
         &  (mag\,arcsec$^{-2}$) & (kpc) & (mag\,arcsec$^{-2}$) & (kpc) \\[0.5ex]
    \hline
    GALEX FUV  & $28.00$ & $18.71$ & --- & ---  \\[+0.5ex]
    GALEX NUV  & $26.75$ & $11.56$ & 28.47 & 7.02  \\[+0.5ex]
    LDT $u$    & $25.70$ & $10.89$ & 24.45 & 2.34 \\[+0.5ex]
    {\sl HERON} $r$ & $24.35$ & $10.60$ & 21.77 & 1.82 \\[+0.5ex]
    IRAC $3.6$    & $22.76$ & $6.13$ & 20.37 & 1.36 \\[+0.5ex]
    VLA H\sc{i} & $0.53^{*}$ & $25.05$ & --- & ---    \\[+0.5ex]    

    \hline\\[-0.5ex]
    \end{tabular}
     \parbox[t]{85mm}{
    \textbf{Notes:}
    $^{*}$ The average H{\sc i} surface density in $M_{\odot}\,\mathrm{pc}^{-2}$ estimated within a circular isophote of 29~mag\,arcsec$^{-2}$ in the GALEX NUV band.\\
    }  
    
\end{table}

\begin{figure}
	\includegraphics[width=0.48\textwidth]{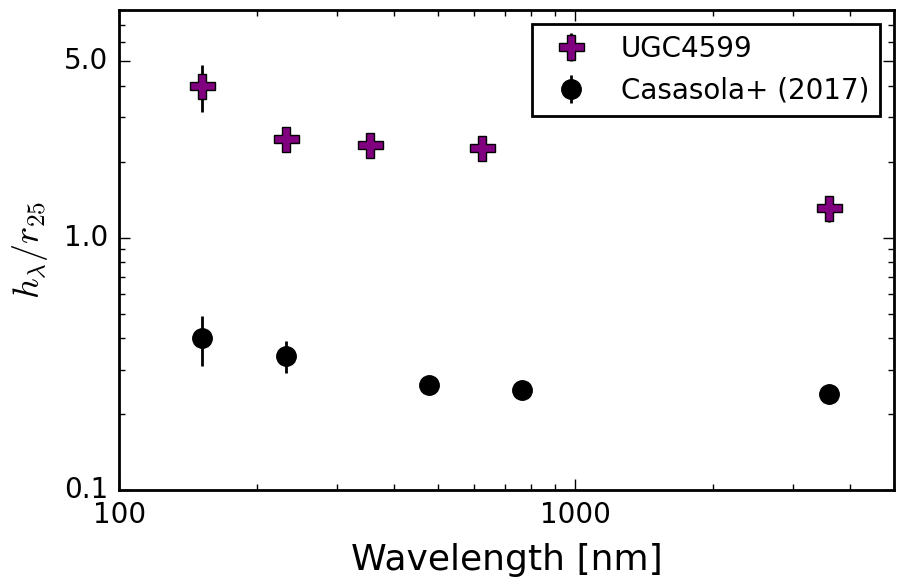}
		\caption{Dependence of the disk scale length normalized by the optical radius from the HyperLeda database \citep{2014A&A...570A..13M} on wavelength. The purple crosses depict UGC\,4599 (see Table~\ref{tab:ProfilesFitting}), whereas the black filled circles correspond to the best-fit results for a sample of 18 nearby face-on spiral galaxies from \citet{2017A&A...605A..18C} (see their table~7).}
	\label{h_lambda}
\end{figure}

As can be seen from Table~\ref{tab:ProfilesFitting}, the disk scale length steadily decreases with wavelength, so the stellar disk at 3.6~$\mu$m is 3 times shorter than the FUV disk (the 3.6~$\mu$m profile, however, only traces the inner galaxy disk). This trend is well-known in the literature and is interpreted by the inside-out formation of the stellar disk \citep[see e.g.,][and references therein]{2007ApJ...658.1006M}. In Fig.~\ref{h_lambda}, we compare this trend with that for regular spiral galaxies from \citet{2017A&A...605A..18C}: the decrease of the scale length for normal spirals appears to be less dramatic ($h_\mathrm{FUV}\sim1.7\,h_\mathrm{3.6}$) than for UGC\,4599. Also, due to the faint nature of the stellar disk in UGC\,4599, the scale length of the galaxy in the visible appears to be 1.5-4 times larger than its optical radius, whereas for normal disks the $h/r_{25}$ ratio is close to 1/4. The extended UV emission, coupled with the enormous H{\sc i} disk ($\sim50$~kpc in radius, see also \citealt{2009A&A...498..407G}), qualify UGC\,4599 as an extended ultraviolet disk (XUV-disk) galaxy \citep{2007ApJS..173..538T}.

As shown by \citet{2017A&A...605A..18C}, for regular spiral galaxies the scale length of the gas profile is roughly equal to the scale length in the FUV band, so  $h_\mathrm{\Sigma gas}/r_{25}=0.40\pm0.07$. For UGC\,4599, the scale length of the neutral hydrogen gas is 1.3 of the FUV scale length. Since the gas mass in LSB galaxies is dominated by atomic gas (\citealt{2009ApJ...696.1834W} and see a recent study by \citealt{2022ApJ...940L..37G} for the CO mass in Malin~1), we can assume that the H{\sc i} gas density profile demonstrates the general distribution of gas in UGC\,4599. Note also that the observed inner depression in the H{\sc i} disk is typical of spiral galaxies \citep{2014MNRAS.441.2159W}.

Note also that, similar to the disk scale length, the effective radius of the bulge in UGC\,4599 drops with wavelength, whereas the bulge effective radius in other galaxies usually increases from blue to red colors \citep{2004A&A...415...63M}.

\subsection{2D photometric decomposition of UGC\,4599}
\label{sec:decomposition}	

\begin{figure}
	\includegraphics[width=0.48\textwidth]{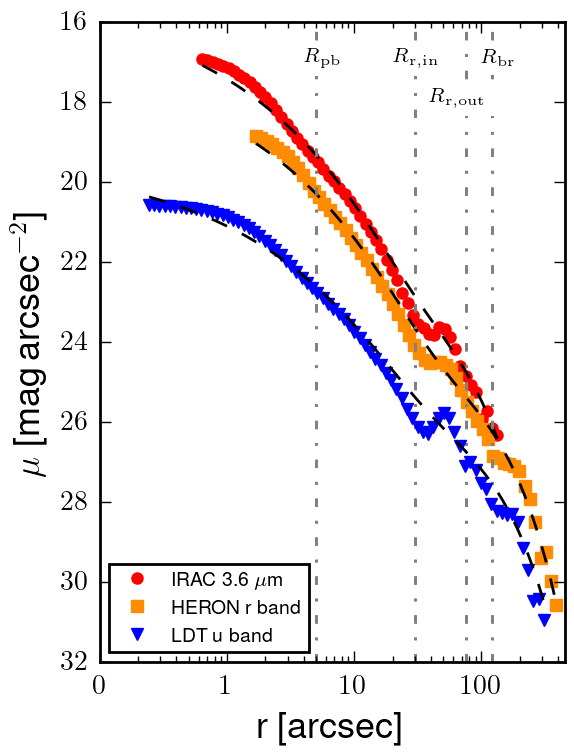}
		\caption{Azimuthally averaged profiles of UGC\,4599 in different wavebands. Dashed lines show regions where the next component begin to dominate.}
	\label{log_profiles}
\end{figure}

To fully characterize the structure of UGC\,4599, we produce a photometric model of the galaxy utilizing its entire 2D image. This is done through the use of the {\tt{IMFIT}} code \citep{2015ApJ...799..226E}, which fits parametric models to point and extended sources in astronomical images. The program supports building simulated galaxies from scratch and fitting the structural parameters of existing galaxies from images. Here we make use of {\tt{IMFIT}} to not only obtain the parameters of the structural components for UGC\,4599, but also to amplify the visible spiral structure in the residuals as shown in Sect.~\ref{sec:spiral_structure}. 

Creating an {\tt{IMFIT}} photometric model for UGC\,4599 implies decomposing it into several structural components. We point out that the core of the galaxy is more complex than was assumed in our simple 1D modeling in Sect.~\ref{sec:profiles}. Fig.~\ref{log_profiles} clearly demonstrates that within a radius of $R_\mathrm{pb}\sim5$~arcsec, our 1D models cannot adequately describe the galaxy profiles suggesting the presence of another, unresolved component in the galaxy core. Therefore, we decided to add this component in our 2D model, along with a S\'ersic bulge (which dominates at $R_\mathrm{pb}<r<R_\mathrm{r,in}$) and a broken-exponential disk dominating at $r>R_\mathrm{r,out}$ with the break radius $R_\mathrm{br}$. Since the spiral arms are very faint and narrow, we mask them out and neglect their presence in our decomposition. We note, however, that we attempted to model the galaxy with an additional two-armed spiral component but our models never converged because the spiral structure in this galaxy is rather complex (see Sect.~\ref{sec:spiral_structure}) and faint, so its weight in the models is too small to be reliably fitted. Also, we mask off the star-forming ring within  $R_\mathrm{in}<r<R_\mathrm{r,out}$ to make our model simpler and better convergeable.

We carried out 2D photometric decomposition for both the {\sl HERON} and IRAC 3.6~$\mu$m images to demonstrate the adequateness of our models in describing the galaxy structure. For the IRAC image, we used a single exponential model because its field of view does not capture the outer disk.

\begin{figure*}
\includegraphics[width=0.46\textwidth]{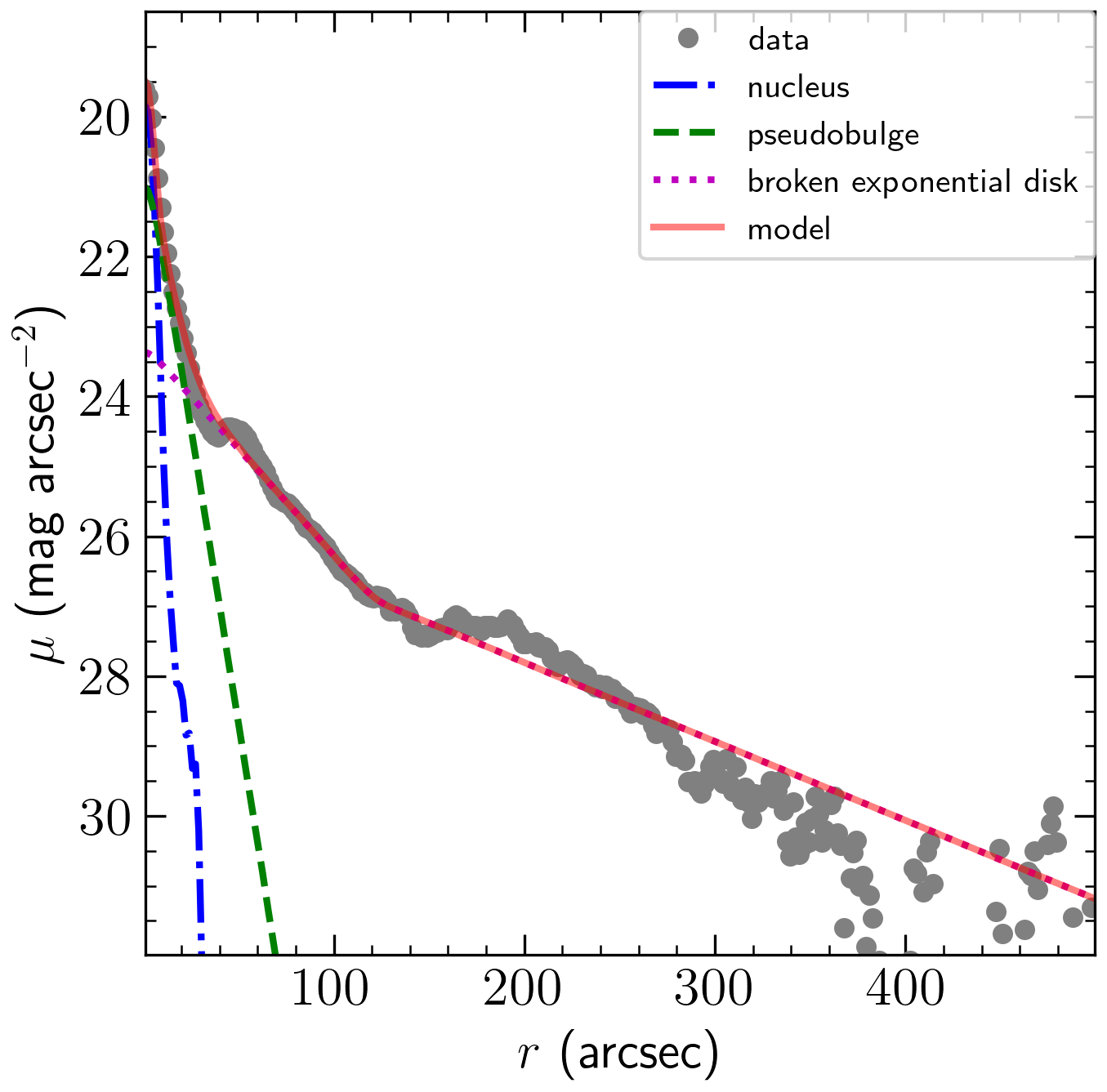}
\includegraphics[width=0.46\textwidth]{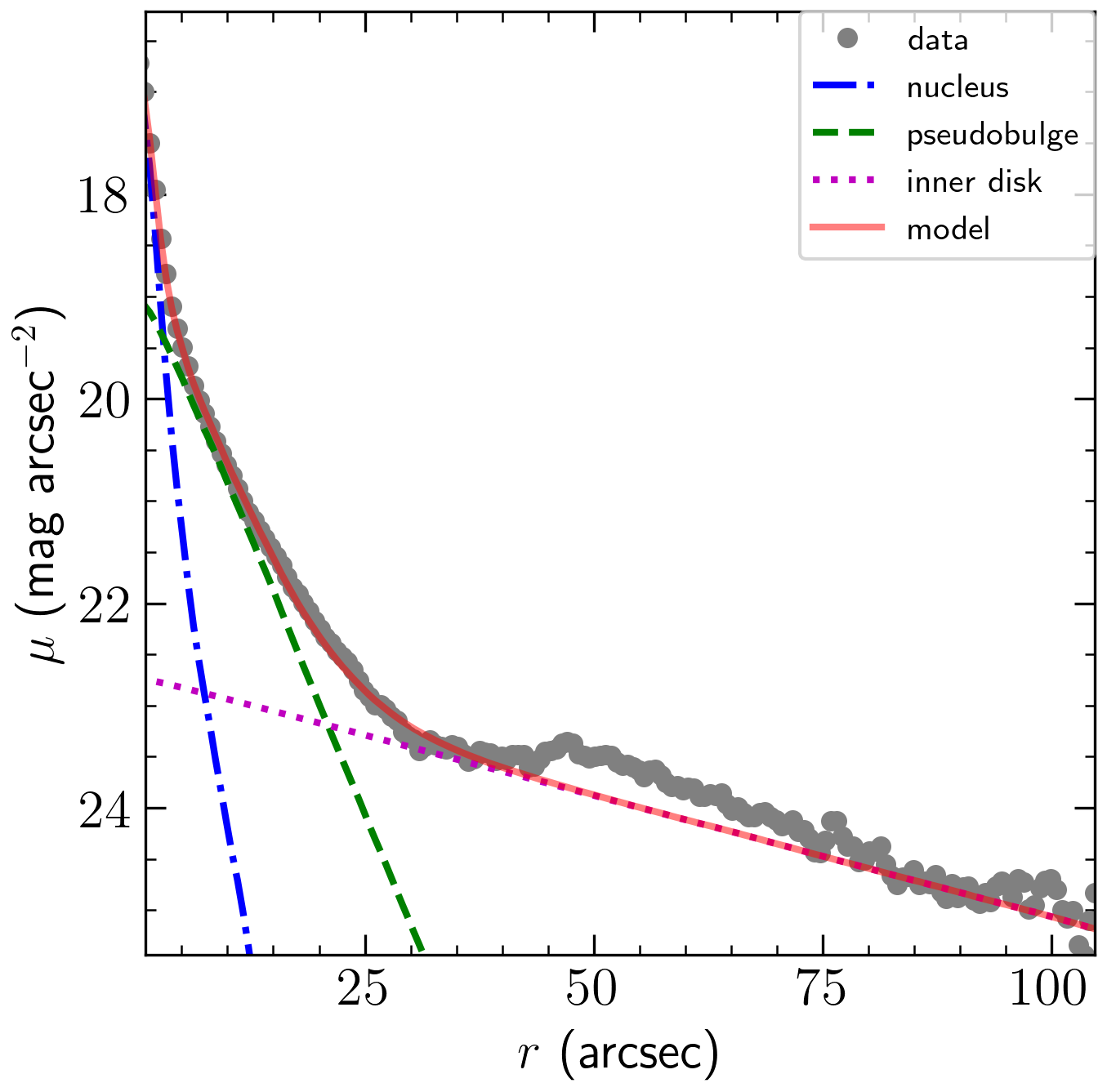}
\caption{Azimuthally-averaged profile of UGC\,4599 for the {\sl HERON} ({\it left}) 
 and IRAC 3.6~$\mu$m ({\it right}) images with the corresponding {\tt{IMFIT}} models superimposed (see Table~\ref{tab:UGC4599_decomp.tab}). Spiral structure and the inner ring was masked in preprocessing before this fit.}
\label{imfit1}
\centering
\end{figure*}

The results of our fitting are presented in Table~\ref{tab:UGC4599_decomp.tab} and illustrated in Fig.~\ref{imfit1}. Our models suggest that the nucleus of the galaxy is indeed characterized by an unresolved source: in the {\sl HERON} we used a PSF, whereas for the IRAC model we employed a S\'ersic function, but the effective radius appeared to be smaller than FWHM/2 (FWHM$_{3.6\mu\mathrm{m}}=2.1$~arcsec), so this component cannot be resolved at 3.6~$\mu$m as well. The galaxy might host an active galactic nucleus (AGN), but its spectrum does not contain the characteristic AGN features. Therefore, we speculate that the nuclear component (within a radius of $\sim200$~pc) may be a nuclear star cluster or any other compact stellar component possibly related to the inner spiral structure which seems to be emerging from the innermost resolved region of a few kpc (see Sect.~\ref{sec:spiral_structure}).

The other inner component, which makes up the bulk of the core's luminosity, is a disk-like structure, possibly a pseudobulge. This assumption is confirmed by 1) the exponential behaviour of the inner galaxy profile outside of $R_\mathrm{pb}\sim5$~arcsec and 2) the inner spiral structure which is detailed in Sect.~\ref{sec:spiral_structure}. \citet{2023A&A...669L..10S} also identify a pseudobulge in their $r$-band profile of UGC\,4599 with the central surface brightness $\mu_\mathrm{0,b}(r)=20.5$~mag\,arcsec$^{-2}$ and $r_\mathrm{e,b}=14.8$~arcsec versus our $\mu_\mathrm{0,b}(r)=20.3$~mag\,arcsec$^{-2}$ and $r_\mathrm{e,b}=10.7$~arcsec for the {\sl HERON} model.

Finally, the inner disk with the central surface brightness $\mu_\mathrm{0,d}=23.2$~mag\,arcsec$^{-2}$ and the scale length $h_\mathrm{in}=35.9$~arcsec is in a relatively good agreement with the results from \citet{2023A&A...669L..10S}, $\mu_\mathrm{0,d}=23.6$~mag\,arcsec$^{-2}$ and $h_\mathrm{in}=49.1$~arcsec, taking into account that our {\sl HERON} model consists of the additional unresolved central component and the outer disk. The outer part of the broken-exponential profile has a very large radial scale $h_\mathrm{out}\approx15$~kpc and low extrapolated central surface brightness $\mu_\mathrm{0,dout}\approx25.5$~mag\,arcsec$^{-2}$. Both values of $\mu_\mathrm{0,d}$ and $\mu_\mathrm{0,dout}$, coupled with the very large disk extent, suggest that UGC\,4599 should be classified as a GLSB galaxy, and the dim inner and outer spiral structure changes its morphology from S0 to S.

\begin{table}
\caption{Results of the {\tt{IMFIT}} fitting for UGC\,4599 for the {\sl HERON} and IRAC 3.6~$\mu$m image. In parentheses we list the {\tt{IMFIT}} functions for describing each galaxy component.}
\label{tab:UGC4599_decomp.tab}
\centering
    \begin{tabular}{lcccc}
    \hline
    \hline\\[-1ex]    
    Component & Parameter &  {\sl HERON} & IRAC & Units \\
              &           &    $r$ band  & 3.6~$\mu$m & \\[0.5ex]
    \hline\\[-0.5ex]
    1. Nucleus  & $M_\mathrm{n}$ & -17.51 & -18.36 & mag \\[+0.5ex]
     ({\it PSF}):& $f_\mathrm{n}$ & 0.107 & 0.140 &  \\[+0.5ex]
    2. Pseudobulge  & $n$  &  1  & 1 &  \\[+0.5ex]
     ({\it S\'ersic})& $\mu_{\mathrm{e,b}}$ & 22.07 & 20.53 & mag\,arcsec$^{-2}$  \\[+0.5ex] 
                     & $r_\mathrm{e,b}$ & 1.67 & 1.36 & kpc \\[+0.5ex]  
                     & $M_\mathrm{b}$ &  $-18.28$ & -19.34 & mag  \\[+0.5ex]
                     & $f_\mathrm{b}$     & $0.219$ & $0.342$    &          \\[+0.5ex]
    3. Disk         & $\mu_{\mathrm{0,d}}$ & $23.16$ & $22.71$ & mag\,arcsec$^{-2}$  \\[+0.5ex]
 ({\it (Broken)} & $h_\mathrm{in}$  &  $5.60$  & $7.36$ &  kpc \\[+0.5ex]
  {\it Exponential}):& $h_\mathrm{out}$  &  $15.03$  & --- &  kpc \\[+0.5ex]
                    & $R_\mathrm{br}$   & 18.98 & --- & kpc \\[+0.5ex]
                     & $M_\mathrm{d}$ &  $-19.50$ & $-20.14$ & mag  \\[+0.5ex]
                     & $f_\mathrm{d}$     & $0.674$ & $0.518$ &           \\[+0.5ex]
    \hline\\[-0.5ex]
    \end{tabular}
     \parbox[t]{80mm}{
    \textbf{Notes:}
    $n$ is the S\'ersic index, $\mu_{\mathrm{e,b}}$ is the effective surface brightness of the S\'ersic component at the effective radius $r_\mathrm{e,b}$;  $\mu_\mathrm{0,d}$ is the disk central surface brightness and $h_\mathrm{in}$ and $h_\mathrm{out}$ are the inner and outer disk scale lengths, respectively. $M$ and $f$ denote the absolute magnitude and the fraction of the component's luminosity, respectively. The surface brightnesses and absolute magnitudes have been corrected for Galactic extinction using \citet{2011ApJ...737..103S} and K-correction using the K-correction calculator \citep{2010MNRAS.405.1409C,2012MNRAS.419.1727C}.}
\end{table}

\subsection{Analysis of the spiral structure in UGC\,4599}
\label{sec:spiral_structure}

\begin{figure}
\includegraphics[width=0.48\textwidth]{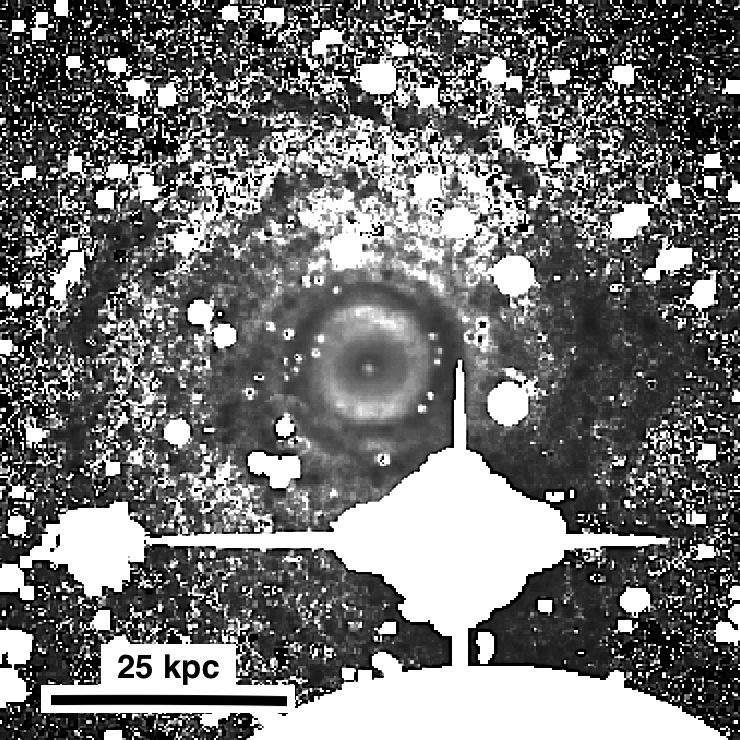}
\caption{Residual {\sl HERON} image of UGC\,4599: (Observed Image-Model)/Observed Image. Dark color displays positive residuals and highlights the spiral structure.}
\label{resid}
\centering
\end{figure}

The faint spiral structure of UGC\,4599 is clearly visible in several passbands (Fig.~\ref{surf_map}). In this section, we exploit the LDT $u$-band, {\sl HERON} and H{\sc i} images to investigate the inner and outer parts of the spiral structure in great detail. The residuals of the {\sl HERON} model in Fig.~\ref{resid} strongly emphasize the spiral arms of the galaxy winding out from the star-forming ring. The illustration of the likely location of the inner part of the spiral structure, which extends inward inside of the ring, is helpful in confirming the pitch angle measurement of the galaxy.

In all UV and optical 1D profiles in Fig.~\ref{azim_profiles}, the ring and spiral arms are well-seen as bumps above the exponential disk model. \citet{2011AJ....142..145G}, based on their deep $B$-band image, also obtained the outer disk profile with faint spiral structure out to $\sim200$~arcsec. In our UV and optical images, we see extended spiral structure in all directions emerging from the tightly wound spiral ring out to $\sim290$~arcsec.

\begin{figure}
		\includegraphics[width=0.48\textwidth]{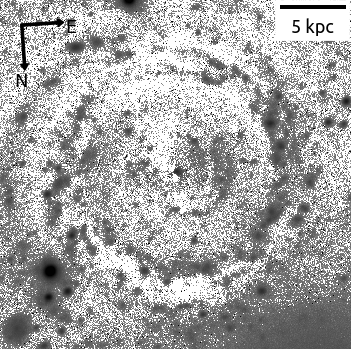}
		\caption{The inner region of UGC\,4599 as shown by the LDT $u$-band image minus an azimuthally-averaged model. The inner spiral (mostly shown in dark), traced by multiple star-forming regions, winds up toward the center.}
		\label{fig:uvimg}
\end{figure}

While present in the GALEX images, the LDT image of UGC\,4599 in the $u$ band displays the faint structure of the galaxy in great detail: the LDT resolution is 1.7~arcsec versus $4.0$~arcsec for GALEX FUV and $5.6$~arcsec for GALEX NUV. The zoomed-in inner part of the $u$-band image in Fig.~\ref{fig:uvimg} clearly shows the inner spiral structure (supported by the {\sl HERON} image, which has a poorer resolution and a much larger pixel scale). The conclusion, made in \citet{2011MNRAS.413.2621F} that the core of UGC\,4599 is surrounded by a detached ring, is thus not confirmed. In the LDT and {\sl HERON} images, we can see one spiral arm emerging from the nucleus. The estimated brightness of this inner spiral structure after subtracting the azimuthally averaged model, obtained in Sect.~\ref{sec:profiles}, is in a range between 27 and 25 mag\,arcsec$^{-2}$ in the $u$ passband. It is important to note that the shape of the inner spiral arm changes with decreasing distance to the center: it becomes less round and less wound.

In addition to the UV and optical images, the column density mapping of H{\sc i} from the VLA confirms the presence of the two-armed spiral structure in H{\sc i}, as described below.

To retrieve the parameters of the spiral arms in UGC\,4599, we employ a method based on an accurate analysis of photometric cuts perpendicular to the arm direction in a galaxy \citep{2020MNRAS.493..390S,2020RAA....20..120M}. Each slice, averaged along some segment of the spiral arm where the curvature of the arm does not change significantly, is fitted (taking into account PSF smearing) with an asymmetric Gaussian function with the `inward' $w_1$ and `outward' $w_2$ half-widths plus some linear trend arising due to the presence of the dominant structural component(s) --- the bulge, the ring, and the disk (see the detailed description of the method in \citealt{2020MNRAS.493..390S}). This allows us to measure the arm width, asymmetry, pitch angle, and the variations of these parameters with radius. With the aid of this method, we are able to determine the characteristics of the spiral structure in UGC\,4599 notwithstanding its low surface brightness nature.

The arms are mainly traced based on the {\tt{IMFIT}} residual image shown in Fig.~\ref{resid}. We also measure clearly visible arcs/parts of the spiral arms in the $u$-band image assuming that the spiral structure behaves in a similar way in the UV and in the optical. Although most segments of the spiral structure are seen well in Fig.~\ref{resid}, there are several places where the spiral arm(s) cannot be continued or the resolution of our images is insufficient to see details. Such places also include the brightest foreground star, which unfortunately outshines the segment where one arm separates into two. There are several places where the arms are located very close to each other, especially in the inner part of the galaxy, so it is difficult to discern the spiral pattern from the ring and the bulge. Using the same approach, we also trace the arms in the H{\sc i} image.

\begin{figure}
	\includegraphics[width=0.45\textwidth]{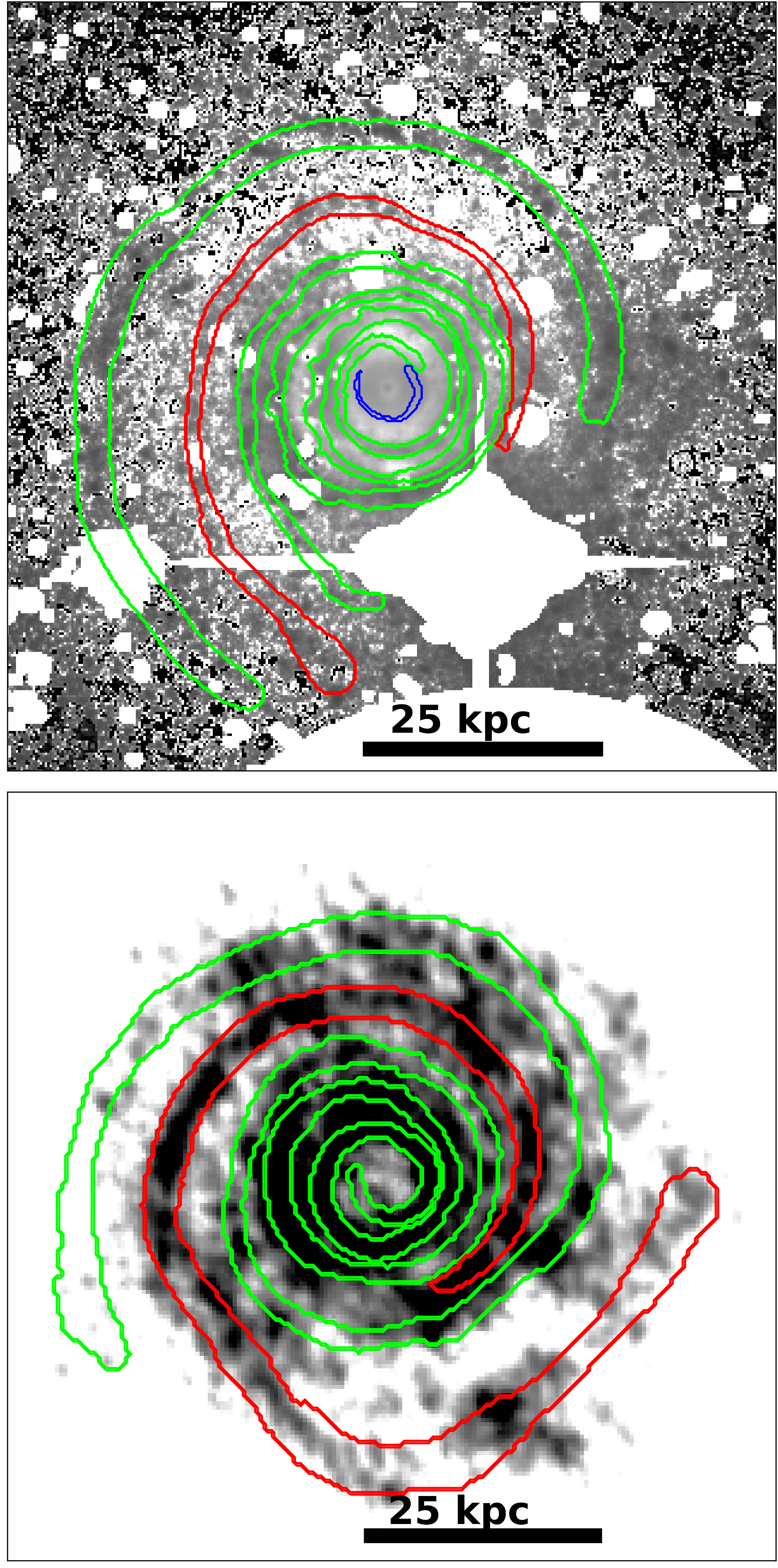}
		\caption{Smoothed masks for the spiral structure in UGC\,4599, based on the photometric cuts made perpendicular to the tangent along the spiral arms. The underlying image in the upper panel displays the {\sl HERON} data residue after the model subtraction; the bottom image shows the H{\sc i} data. The color curves outline different spiral arms. The blue arc corresponds to the LDT $u$-band data.}
	\label{cuts_mask}
\end{figure}

By means of our method, we successfully traced the spiral structure from the inner region out to a radius of 45~kpc. The resulting spiral structure, mapped individually for the LDT $u$-band, {\sl HERON}, and H{\sc i} data, is presented in Fig.~\ref{cuts_mask}. It is easy to see that although the masks of the spiral arms, highlighted by the stars and gas, look similar in general, they are slightly different in details. This fact can be due to the different resolution of the images used and the slightly different locations of the star-forming regions in the optical and UV for the spirals and the atomic gas.

\begin{figure}
	\includegraphics[width=0.49\textwidth]{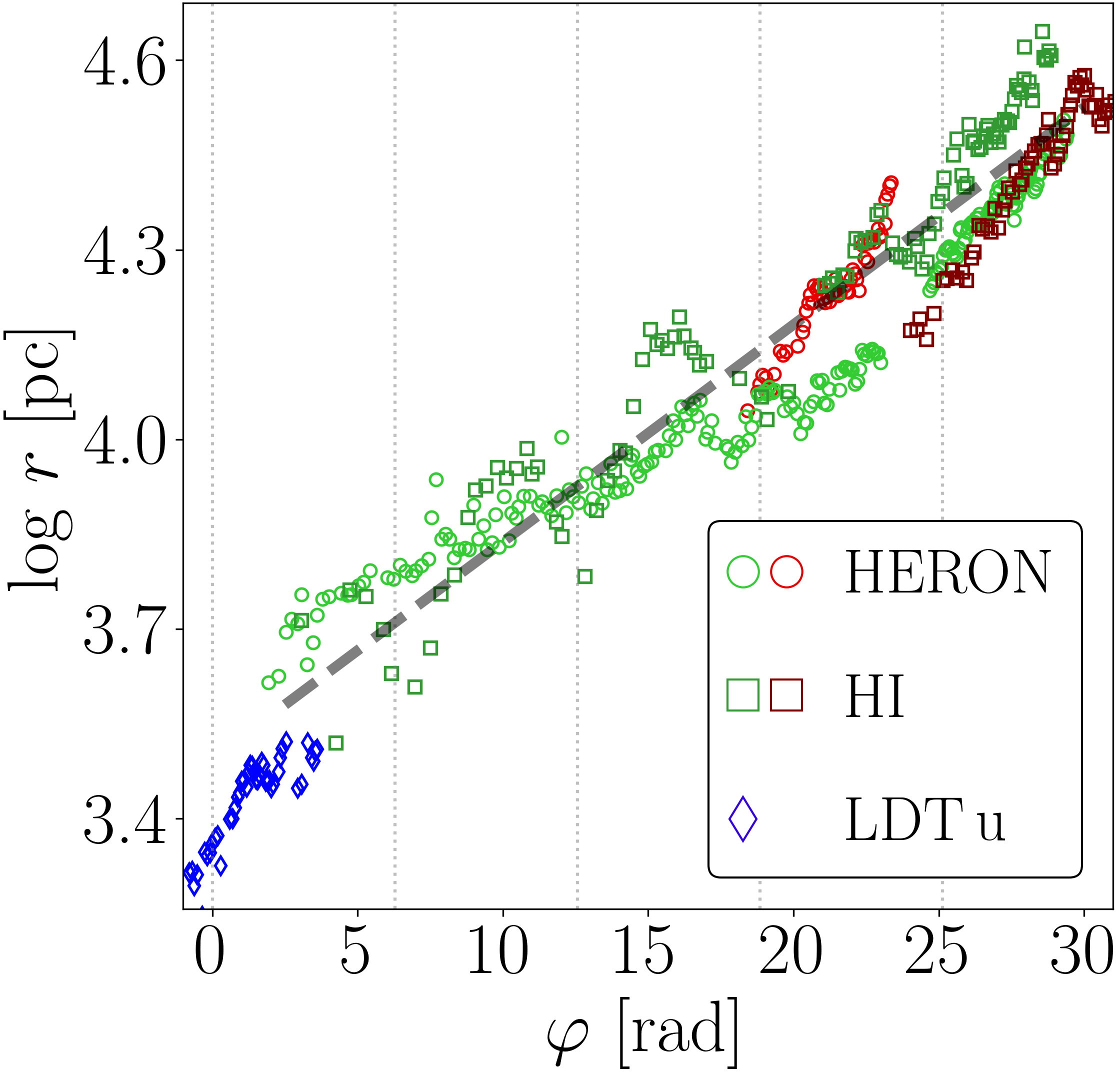}
		\caption{Pitch angle estimation for the spiral arms in UGC\,4599 in the log-polar coordinate system based on the LDT $u$-band (inner region), {\sl HERON} $r$-band, and H{\sc i} data. Red symbols depict the split spiral arm, as shown with the same color in Fig.~\ref{cuts_mask}. Blue symbols depict the inner spiral arm traced in the LDT $u$-band image. The grey line represents the best linear fit to the {\sl HERON} data with a pitch angle of $4.5\degr$.}
	\label{cut_spiral_polar}
\end{figure}

As was noted earlier, Fig.~\ref{fig:uvimg} demonstrates the presence of one inner arm emerging from the pseudobulge and winding out to form the ring. The inner spiral arm, depicted by blue color in Figs.~\ref{cuts_mask} and \ref{cut_spiral_polar}, has a pitch angle of $7.0\degr \pm 0.4\degr$. In the place blanketed by the bright star, we find that the arm splits into two distinct outer arms. The fact that the galaxy has two outer spiral arms is also confirmed by the H{\sc i} image, which clearly displays the two arm ends at the periphery of the galaxy. 

The peaks in the polar coordinates presented in Fig.~\ref{cut_spiral_polar} also demonstrate a complex behavior of the spiral arms. We see that the H{\sc I} and optical data coincide well across the azimuthal angle in most of the places which strengthens our results. Fitting the entire data, we derive a pitch angle of approximately 4.5\degr for both distributions. At the same time, we can see in Fig.~\ref{cut_spiral_polar} that one of the split arms, marked by the red color, acts differently after the split (bifurcation) point. This arm demonstrates a different slope, agreed between the H{\sc i} and stellar data, resulting in a twice as large pitch angle of 8.2\degr. This can be additional evidence in favor of the two-armed model. The average pitch angle between the two spiral arms is $P=6.4\degr$, which we will use for comparison with other spiral galaxies in Sect.~\ref{sec:wave_theory}. 

\section{Discussion}
\label{sec:discussion}

We now proceed to a quantitative comparison of UGC\,4599 to other galaxies to elucidate its nature and discuss various scenarios of its formation.

\subsection{Galaxy scaling relations}
\label{sec:scaling_relations}

It is interesting to see how UGC\,4599 compares with other galaxies (including other GLSB galaxies) on standard galaxy scaling relations. For that, we also performed a photometric decomposition of Hoag's Object (see Appendix~\ref{sec:hoag}) to demonstrate that UGC\,4599 and Hoag's Object are not only different morphologically (UGC\,4599 is a GLSB, ring spiral galaxy, whereas Hoag's Object is an elliptical galaxy surrounded by a star-forming ring), but also have different locations on the scaling relations.

For comparison purposes, we display the results of both 1D and 2D photometric decomposition  for the {\sl HERON} data detailed in Sect.~\ref{sec:analysis}.

\begin{figure*}
        \includegraphics[height=0.47\textwidth]{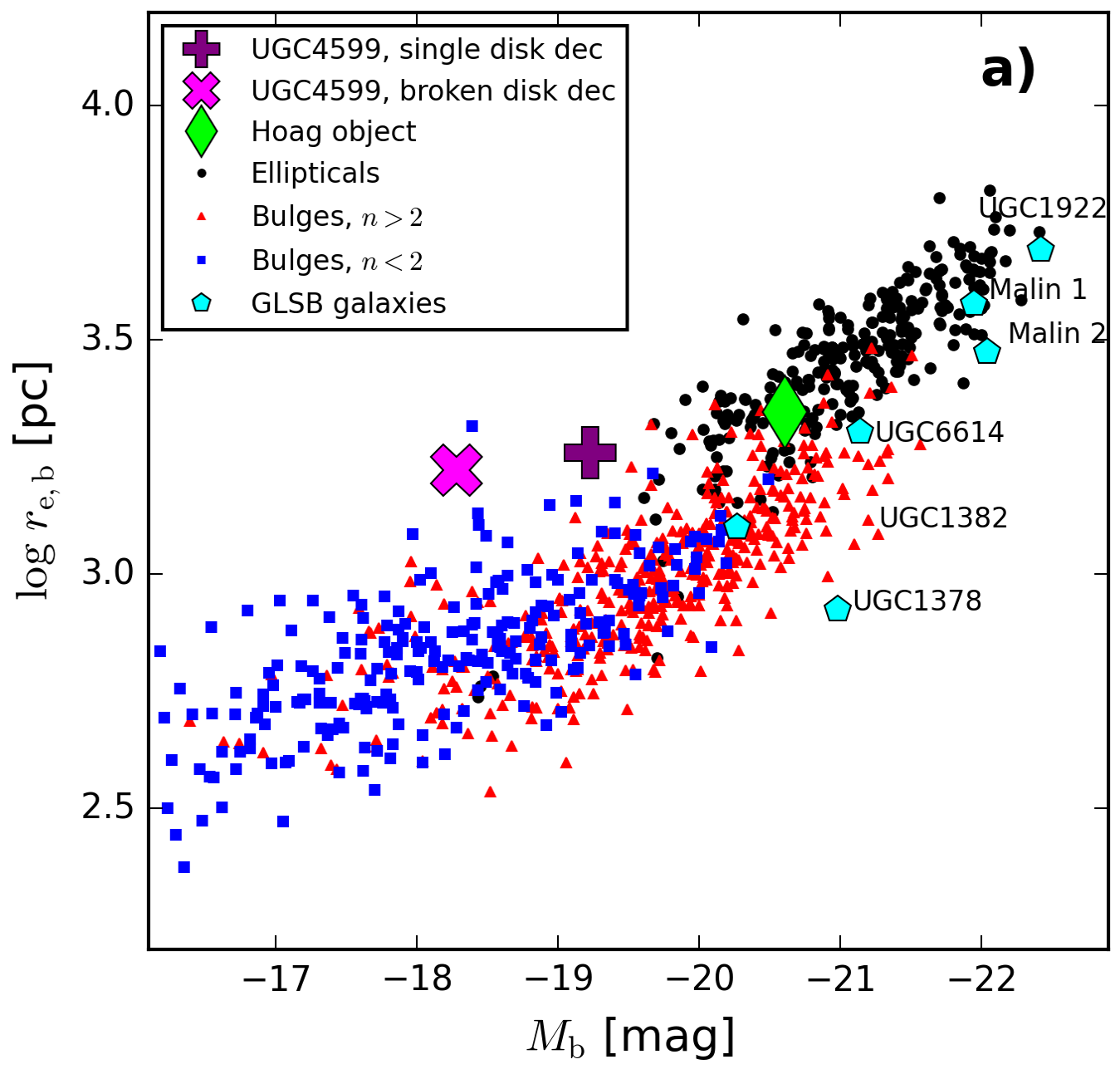}
	\includegraphics[height=0.47\textwidth]{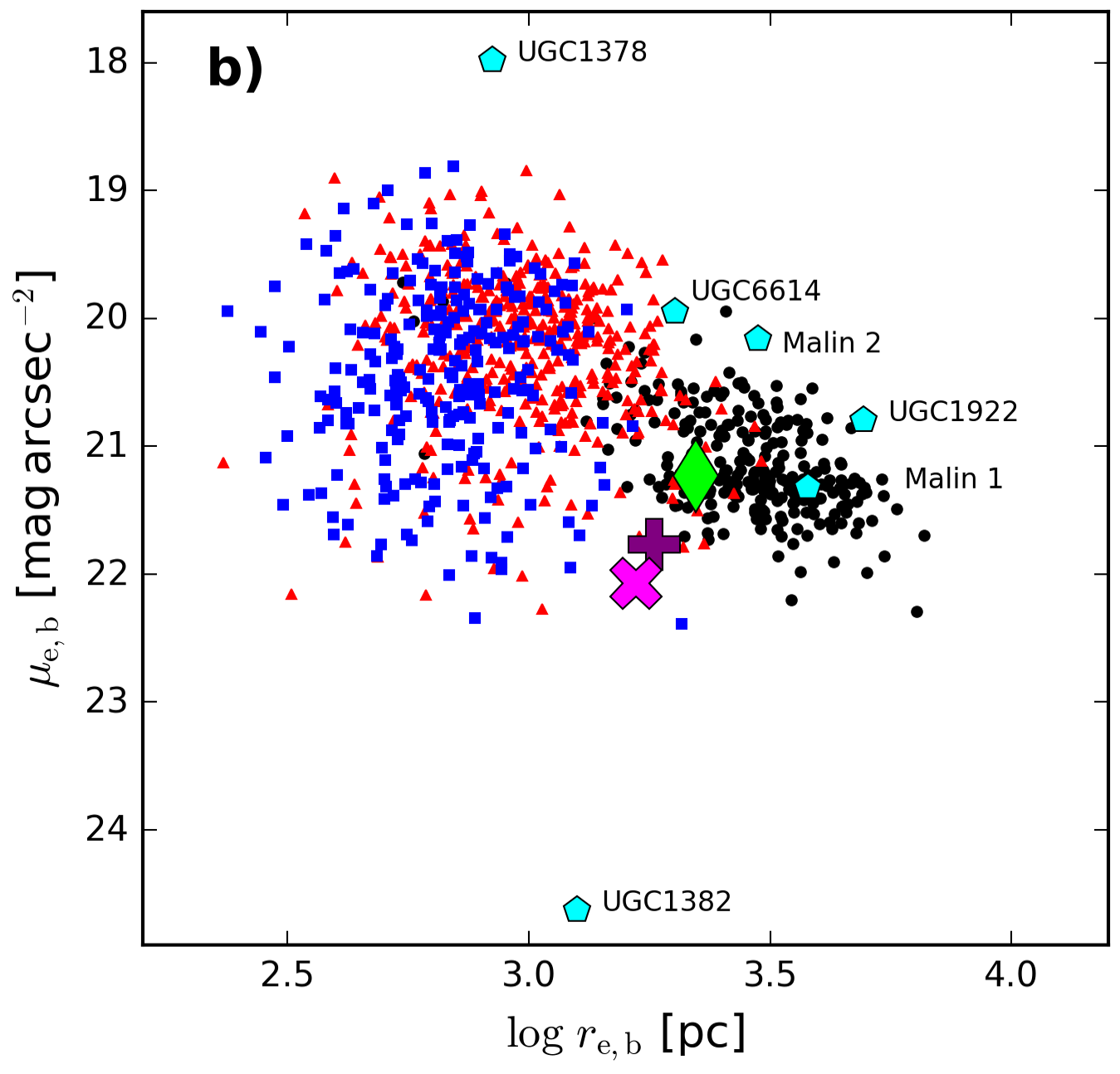}
        \includegraphics[height=0.478\textwidth]{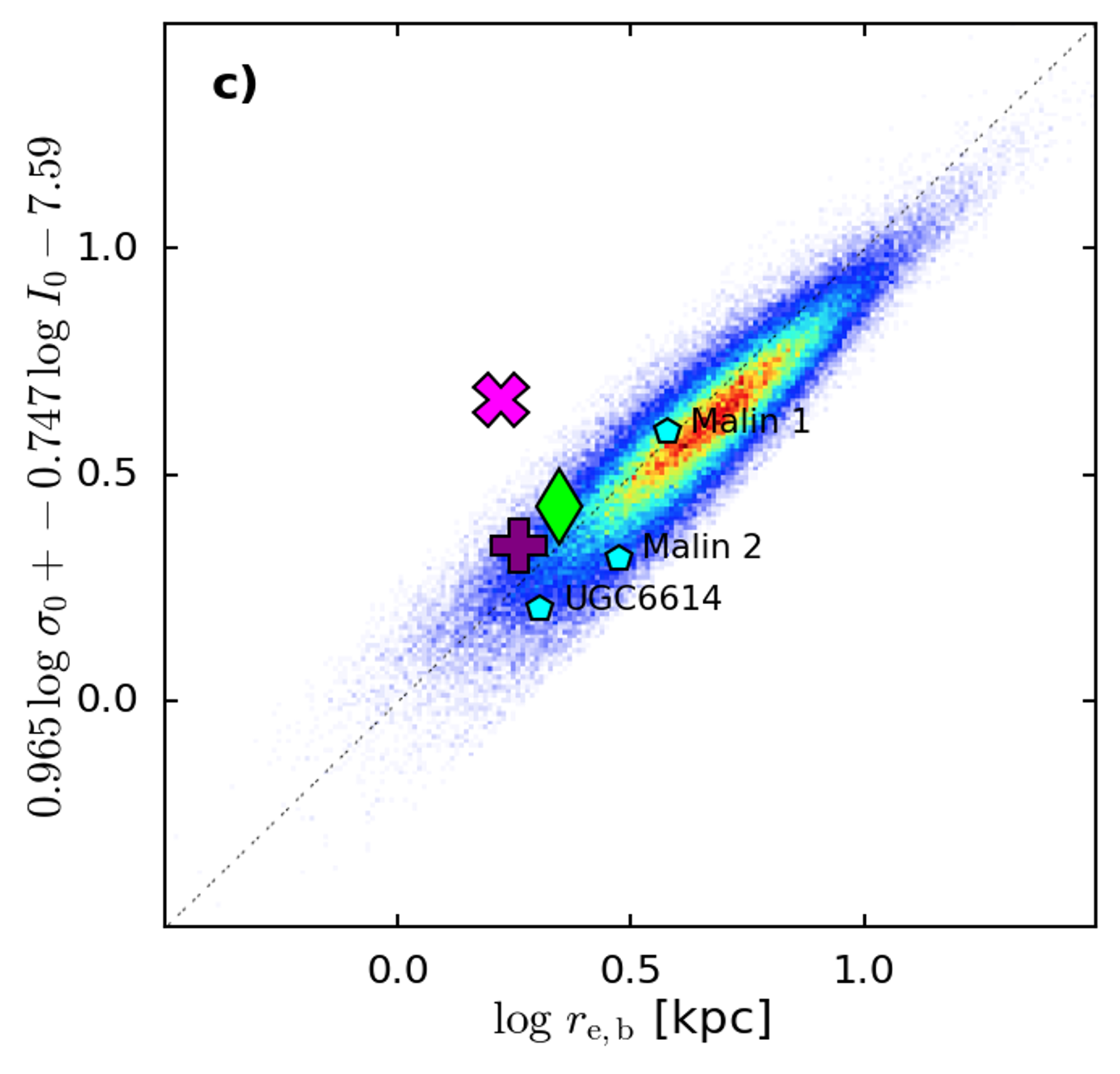}
        \includegraphics[height=0.47\textwidth]{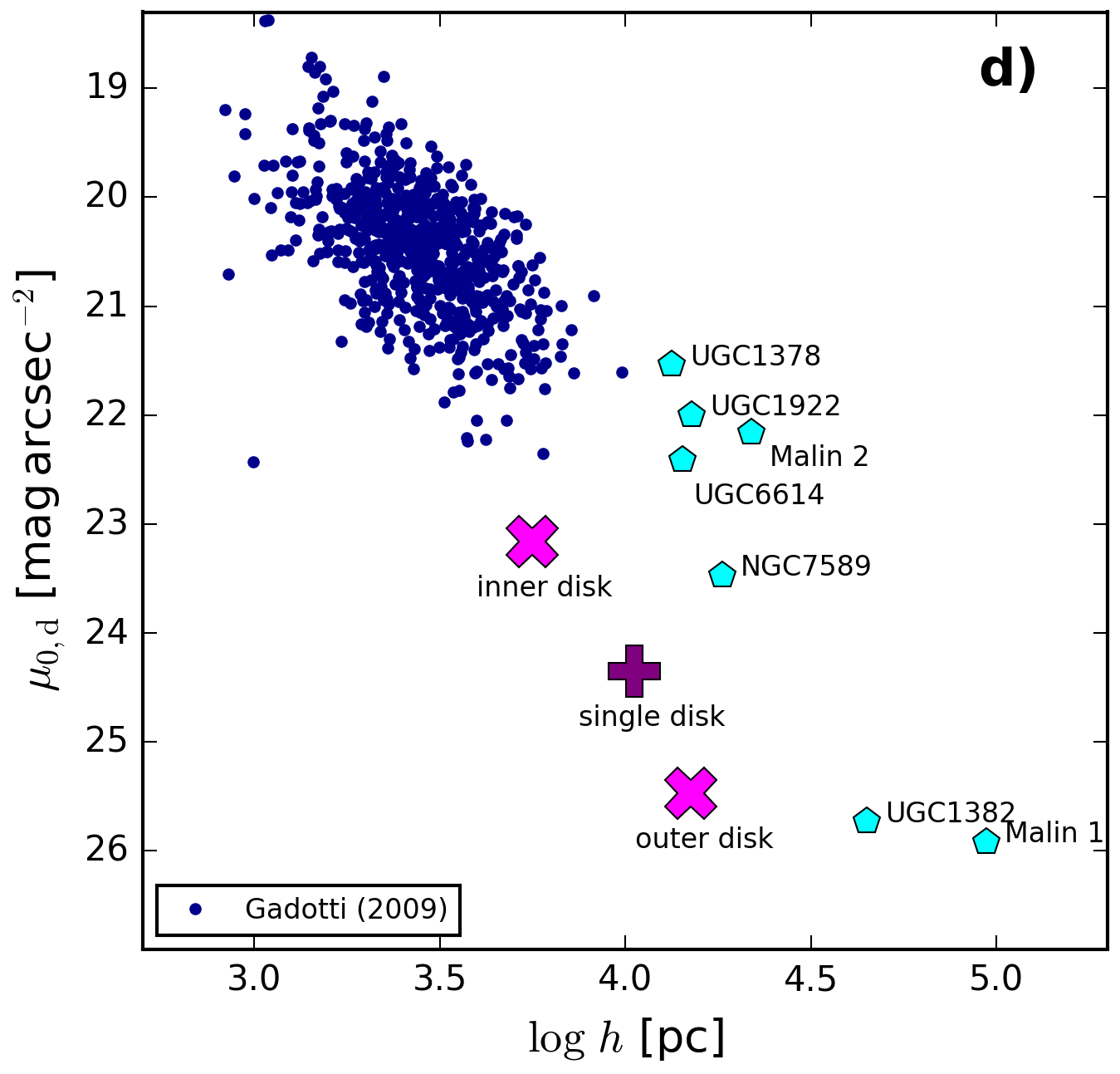}
		\caption{{\it a)} The effective radius vs. bulge luminosity relation for the samples of galaxies from \citet{2009MNRAS.393.1531G} and GLSB galaxies from \citet{2021MNRAS.503..830S}. {\it b)} Kormendy relation for the same samples as in {\it a)}. {\it c)} The fundamental plane, with coefficients taken from \citet{2013A&A...557A..21S} for de Vaucouleurs' models of elliptical galaxies (the scatter plot is adopted from their study, see their fig.~B.51). The renormalised surface brightness $I_0$ is related to the mean effective surface brightness $\langle \mu_\mathrm{e,b} \rangle$ through $\log I_0 = -\langle \mu_\mathrm{e,b} \rangle/2.5$. {\it d)} $\mu_\mathrm{0,d}$--$h$ relation for the stellar disks from \citet{2009MNRAS.393.1531G}. The structural parameters for the GLSB galaxies are taken from the following studies: Malin~1 \citep{1997AJ....114.1858P}, Malin~2 \citep{2014MNRAS.437.3072K}, NGC\,7589 \citep{2021MNRAS.503..830S}, UGC\,1378 \citep{2019MNRAS.489.4669S}, UGC\,1382 \citep{2016ApJ...826..210H}, UGC\,1922 \citep{2018MNRAS.481.3534S}, and UGC\,6614 \citep{2021MNRAS.503..830S}. }
	\label{fig:cors}
\end{figure*}

In Fig.~\ref{fig:cors}, plots {\it a} and {\it b}, we show the location of UGC\,4599 on the size--luminosity and Kormendy relations for bulges and elliptical galaxies. We use two samples of galaxies: regular early- and late-type galaxies from \citet{2009MNRAS.393.1531G} 
and GLSB galaxies from \citet{2021MNRAS.503..830S}. As one can see, the bulge of UGC\,4599 strongly deviates from pseudo- and classical bulges, as well as from elliptical galaxies: it has an effective radius typical of classical bulges, but a very low surface brightness, more typical of pseudobulges. Hoag's Object, however, belongs to the locus of elliptical galaxies and does not deviate from the general trends. In contrast to UGC\,4599, the bulges in the displayed GLSB galaxies are, on average, more luminous and have a larger effective radius than the bulges in regular spiral galaxies, deviating from the Kormendy relation for the classical bulges, but following the general trend on the size-luminosity relation.

In Fig.~\ref{fig:cors}{\it c}, we depict the fundamental plane, which links the three global parameters of elliptical galaxies \citep{1987ApJ...313...59D} and bulges \citep{2002MNRAS.335..741F}: the central velocity dispersion (in km/s), the effective radius (in kpc), and the mean surface brightness within the effective radius (in mag\,arcsec$^{-2}$). As one can see, the pseudobulge of UGC\,4599 from our 2D model strongly deviates from the general trend, whereas Hoag's Object and GLSB galaxies generally follow the fundamental plane.

Finally, on the $\mu_\mathrm{0,d}$--$h$ relation (Fig.~\ref{fig:cors}, {\it d}), the disk of UGC\,4599 appears to be extremely faint and large, so it is an obvious outlier from the general trend for the regular disks, whereas the other GLSB galaxies follow this relation but at much lower surface brightnesses and disk scale lengths. Hoag's Object is not present in the plot because it does not have a stellar disk.

Interestingly, the position of UGC\,4599 on the baryonic Tully-Fisher relation \citep{1977A&A....54..661T,1995ApJ...438...72S,2015ApJ...802...18M}, which connects the total mass of stars and gas with the maximum galaxy rotation velocity, follows the general trend for disk galaxies obtained in \citet{2019MNRAS.484.3267L}, with $\log\,M_\mathrm{bar}/M_{\sun}=10.49$ (see Sect.~\ref{sec:star_formation}) and $V_\mathrm{rot}=170.4$~km/s (computed using the $W_{20}=180.4$~km/s emission line width taken from \citealt{2010PhDT.......102D} and inclination angle $i=27\degr$ from HyperLeda). This is another evidence that UGC\,4599 is not affected by strong starbursts or mergers and its disk is not overheated. Moreover, this galaxy should likely be superthin, according to, for example, \citet{2017MNRAS.465.3784B}. 

In general, we can conclude that UGC\,4599 is a very unusual galaxy considering the galaxy scaling relations. The parameters of its bulge and disk differ from those in regular spiral galaxies. The core of Hoag's Object, however, is consistent with a typical elliptical galaxy and does not deviate from the standard scaling relations for the ellipticals. Our sample of GLSB galaxies is too small to make a fair comparison of UGC\,4599 with these galaxies, but on all scaling relations under consideration this GLSB galaxy has a very different location.

\subsection{Spiral structure in UGC\,4599}
\label{sec:wave_theory}

We can also try to characterize UGC\,4599 in the context of the spiral structure's properties as compared to those in other spiral galaxies. The spiral arms of UGC\,4599 are 
more tightly wound than, on average, observed in other spirals (for example, consider the distribution presented in figure~11 in \citealt{2020MNRAS.493..390S}), although there is a tendency for early-type spirals to exhibit lower pitch angles than measured in late-type spirals (see figure~10~in \citealt{2020ApJ...900..150Y}). One of the interesting details of the spiral structure in UGC\,4599 is the large maximum azimuthal angle (the number of turns of the spiral arms) --- almost $10\pi$~rad that is not typical of regular spiral galaxies \citep{2020MNRAS.493..390S}.

We can also compare another interesting parameter of spiral structure, the width $w = w_1+w_2$ of the spiral arms and how it changes with galactocentric radius $r$ in the form $w \propto a\,r$. On average, for both the H{\sc i} and {\sl HERON} images, the width of the spiral arms increases with radius, similar to what was measured for most of the galaxies in \citet{2020MNRAS.493..390S}, but with a relatively small slope of $a=0.03$ compared to the mean $0.14$ for the \citet{2020MNRAS.493..390S} sample. The width of the spiral arms for the H{\sc i} data are, on average, 1.5 times larger than in the optical, and there is no significant difference between the widths of the two outer arms after the bifurcation point. In contrast, the {\sl HERON} image clearly shows that the arm, which is colour-coded with red in Fig.~\ref{cuts_mask}, is thinner than the other outer arm. This difference amounts, on average, to 40\% according to our analysis. The inner spiral is approximately twice as narrow as the outer arms in both the H{\sc i} and optical images. Normalized by the optical radius $r_{25}$, the average width of the spirals in the optical $(w_1+w_2)/r_{25}$ is close to 0.1, which is a typical value for the sample considered in \citet{2020MNRAS.493..390S} (see their figure~16 and figure~5 in \citealt{2020RAA....20..120M}). The width asymmetry $A=(w_2-w_1)/w_2$ of the arms in UGC\,4599 is close to zero ($A=0.03$ for the H{\sc i} and 0.04 for the {\sl HERON} data) which indicates that, on average, the inward side $w_1$ of the arms in UGC\,4599 is approximately the same as the outward side $w_2$. As shown in \citet{2020MNRAS.493..390S}, most spiral galaxies have a positive average width asymmetry $A>0.1-0.2$ (see their figure~18) consistent with the Modal Density Wave Theory \citep{1969ApJ...155..721L}: inside of the co-rotation radius, the shock is formed behind the spiral arm and, thus, $w_2$ should be larger than $w_1$ \citep{2012A&A...542A..39G}. Therefore, we can suggest that this mechanism may be less important for inducing the extremely faint spiral arms in UGC\,4599. 

A correlation between the mass of galactic supermassive black holes and spiral galaxy pitch angle has been obtained in many studies \citep{2008ApJ...678L..93S, 2013ApJ...769..132B, 2017MNRAS.471.2187D, 2015ApJ...802L..13D}. This so-called M-P relation is advantageous for estimating the SMBH mass, especially when a well defined velocity dispersion for the galaxy is absent or when other methods are not available. The M-P relation has also been shown to have less scatter than other methods when applied to spiral galaxies \citep{2017MNRAS.471.2187D}. However, \citet{2019A&A...631A..94D} did not find any strong correlation between mean pitch angle and SMBH mass (estimated from central stellar velocity dispersion) for nearby spiral galaxies from the Spitzer Survey of Stellar Structure in Galaxies (S$^4$G, \citealt{2010PASP..122.1397S}).

SMBH masses, calculated for UGC\,4599 using different scaling relations, are tabulated in Table~\ref{BlackholeMass}. Using equation (8) from \citet{2017MNRAS.471.2187D} (if we assume that the M-P relation holds true for UGC\,4599), we arrive at an estimated SMBH mass for this galaxy of $\log(M/M_{\sun})=8.48 \pm 0.16$, corresponding to a black hole mass of $3.01\times10^8\,M_{\sun}$. This places UGC\,4599 at the very high end of the SMBH mass distribution for spiral galaxies and the lower end for elliptical galaxies (for late-type SMBH mass functions, see \citealt{2014ApJ...789..124D}; for a thorough review of black hole mass functions see \citealt{2012AdAst2012E...7K}). The $M_\mathrm{BH}-n$ log quadratic relation \citep{GrahamDriver2007} yields, however, a three orders of magnitude less-massive SMBH ($3.20 \times 10^5\,M_\odot$) assuming an exponential pseudobulge, while the $M_\mathrm{BH}-\sigma$ relation \citep{Nandini(2019)} gives a mass of $1.13 \times 10^6\,M_\odot$ (adopting the dispersion velocity value $86.67 \pm  2.78\,\mathrm{km/s}$ from SDSS DR12). \citet{Graham2007} summarizes and compares work done by multiple authors on black hole mass spheroid luminosity scaling relations. Since we have derived the properties of the galaxy components using an $r$-band calibrated image, we use the $\log(M_\mathrm{BH}/M_\odot )$=$(-0.38\pm0.04)(M_R + 21)$ + $(8.12\pm0.08)$ relation from \citet{Graham2007}, employing the $R$-band bulge absolute magnitude -19.34 mag (see Table~\ref{tab:UGC4599_decomp.tab}) and find $M_\mathrm{BH}=1.2\times10^7\,M_{\sun}$. 

As one can see, these estimates of the SMBH mass for UGC\,4599 are totally inconsistent among themselves which also proves the peculiar nature of this galaxy.

\begin{table}
\caption{ Black hole mass estimates derived from various scaling relations }
\label{BlackholeMass}
\centering
    \begin{tabular}{cccc}
    \hline
    \hline\\[-1ex]    
    Relation &  Reference & $\log(M_\mathrm{BH}/M_\odot)$ & $M_\mathrm{BH}(M_\odot)$  \\[0.5ex]
    \hline\\[-0.5ex]
    
    $M_\mathrm{BH}-P$ & [1] & $8.48 \pm 0.16$ & $3.01 \times 10^8$   \\[+0.5ex]
    $M_\mathrm{BH}-\sigma$ & [2] & $6.05 \pm 0.14$ & $1.13 \times 10^6$   \\[+0.5ex]
    $M_\mathrm{BH}-n$ & [3] & $5.51 \pm 0.30$& $3.20 \times 10^5$ \\[+0.5ex]
    $M_\mathrm{BH}-L$ & [4] & $7.09 \pm 0.14$ & $1.20 \times 10^7$  \\[+0.5ex]

    \hline\\[-0.5ex]
    \end{tabular}
     \parbox[t]{85mm}{
    \textbf{Notes:}\\
    References: [1] = \cite{2017MNRAS.471.2187D}, [2] = \cite{Nandini(2019)}, [3] = \cite{GrahamDriver2007}, [4] = \cite{Graham2007}. We used the log-quadratic $M_\mathrm{BH}-n$ relation (7) from \citet{Graham2007} employing the S\'ersic index $n = 1$ (see Table 2). For the $M_\mathrm{BH}-L$ relation we used the $\log(M_\mathrm{BH}/M_\odot )=( -0.38\pm0.04)(M_R + 21) + (8.12\pm0.08)$ $R$-band bulge absolute magnitude scaling relation.}

\end{table}

\subsection{Star Formation Rate of UGC\,4599}
\label{sec:star_formation}

The NUV band image from GALEX allows us to infer the star formation rate of the `disk+spirals' in UGC\,4599 as compared to the SFR in the entire galaxy. We do not use the FUV-band image because its photometric depth is poorer than that of the NUV-band image. Also, it is of great importance to compare UGC\,4599 with other galaxies, for example, with galaxies from the ALFALFA-SDSS Catalog \citep{2020AJ....160..271D}, which present stellar masses, star formation rates, and H{\sc i} masses estimated using the same recipes as in this paper.  

The total stellar mass of UGC\,4599 was calculated using the {\it Spitzer} 3.6~$\mu$m data by means of conversion to stellar mass via the $M_{*}/L$ value of \citet{2012AJ....143..139E}. The SFR and the SFR surface density were computed using the GALEX NUV flux and an SFR calibration from \citet{2012ARA&A..50..531K} (their table~1, see also \citealt{2011ApJ...737...67M,2011ApJ...741..124H}):

\begin{equation} \label{eq:1}
\log \mathrm{SFR_{NUV}} = \log \nu L_{\nu} - 43.17\,,
\end{equation}

where $\nu L_{\nu}$ is the NUV spectral energy density uncorrected for the 22~$\mu$m emission. This correction was not done since the galaxy has no detection in the WISE W4 waveband and there are no {\it Spitzer} MIPS observations. This method assumes a constant star formation rate, solar metallicity, and \citet{1955ApJ...121..161S} stellar initial mass function (IMF). We do not account for dust reddening or extinction since i) UGC\,4599 is near to face-on, 2) as shown by \citet{2009ApJ...696.1834W}, the ratio of far-infrared to UV flux for low surface brightness galaxies is significantly less than unity, implying low internal UV attenuation and dust content.   

We estimate the SFR and the surface density of the SFR within both a circular aperture with radius 350~arcsec (the extent of the galaxy in the UV and optical images under study) and an isophote of 29 mag\,arcsec$^{-2}$ in the NUV band to compare our results with those from \citet{2009ApJ...696.1834W} for 19 low surface brightness galaxies. We also provide the total mass and the surface density of the neutral hydrogen gas based on the VLA map. For computing both the stellar mass and SFR of the {\it disk}, we use the disk parameters retrieved in Sect.~\ref{sec:profiles}. The measured quantities are listed in Table~\ref{tab:MassParameters}.


\begin{table}
\caption{Additional characteristics of UGC\,4599 estimated in the present study.}
\label{tab:MassParameters}
\centering
    \begin{tabular}{lcc}
    \hline
    \hline\\[-1ex]    
    Quantity     &  Galaxy & Disk  \\[0.5ex]
    \hline\\[-0.5ex]
    
    $\log M_{*}$ ($M_{\odot}$)  & 10.31 & 10.04  \\[+0.5ex]
    log\,SFR ($M_{\odot}\,$yr$^{-1}$)  & -0.64 & -0.75 \\[+0.5ex]
    log\,SFR$_{29}$ ($M_{\odot}\,$yr$^{-1}$)  & -0.83 & -0.92 \\[+0.5ex]
    $\log \Sigma_\mathrm{SFR}$ ($M_{\odot}\,$yr$^{-1}\,\mathrm{kpc}^{-2}$) & -4.65 & -4.72\\
    $\log \Sigma_\mathrm{SFR,29}$ ($M_{\odot}\,$yr$^{-1}\,\mathrm{kpc}^{-2}$) & -3.80 & -3.89 \\
    $\log M_\mathrm{H{\sc I}}$ ($M_{\odot}$)  & 10.03 & ---  \\[+0.5ex]
    $\log \Sigma_\mathrm{{H{\sc I}}}$ ($M_{\odot}\,\mathrm{pc}^{-2}$) & 0.12 & --- \\
    \hline\\[-0.5ex]
    \end{tabular}
\end{table}

We note here that although the SFR surface density is highest in the pseudobulge and lowest in the spiral arms, 77\% of the total star formation of the galaxy is attributed to the faint but extended disk and the embedded spiral structure. The central body with the ring contribute roughly 23\%. This can be explained by the fact that the diffuse star formation in the outer disk and spiral arms cover a larger area than the core and the ring and, thus, make a greater contribution to the total SFR. We can draw the same conclusion for the stellar mass: despite the very low luminosity of the disk at 3.6~$\mu$m, the disk-to-total mass ratio 0.52 is rather high because of the large extent of the disk.

After describing the star formation in UGC\,4599, we turn to
comparing it to other LSB galaxies in terms of SFR, average $\Sigma_\mathrm{SFR}$, and the gas mass and surface densities. The comparison with 18 LSB galaxies from \citet{2009ApJ...696.1834W} shows that UGC\,4599 has a very typical SFR, $\Sigma_\mathrm{SFR}$, and $ \Sigma_\mathrm{{H{\sc I}}}$. The location of this galaxy in the SFR surface density vs. gas surface density is in good agreement with with the position of other LSB galaxies from \citet{2009ApJ...696.1834W} and GLSB galaxies from \citet{2021MNRAS.503..830S} (see their figure~16). 

\begin{figure*}
	\includegraphics[width=0.49\textwidth]{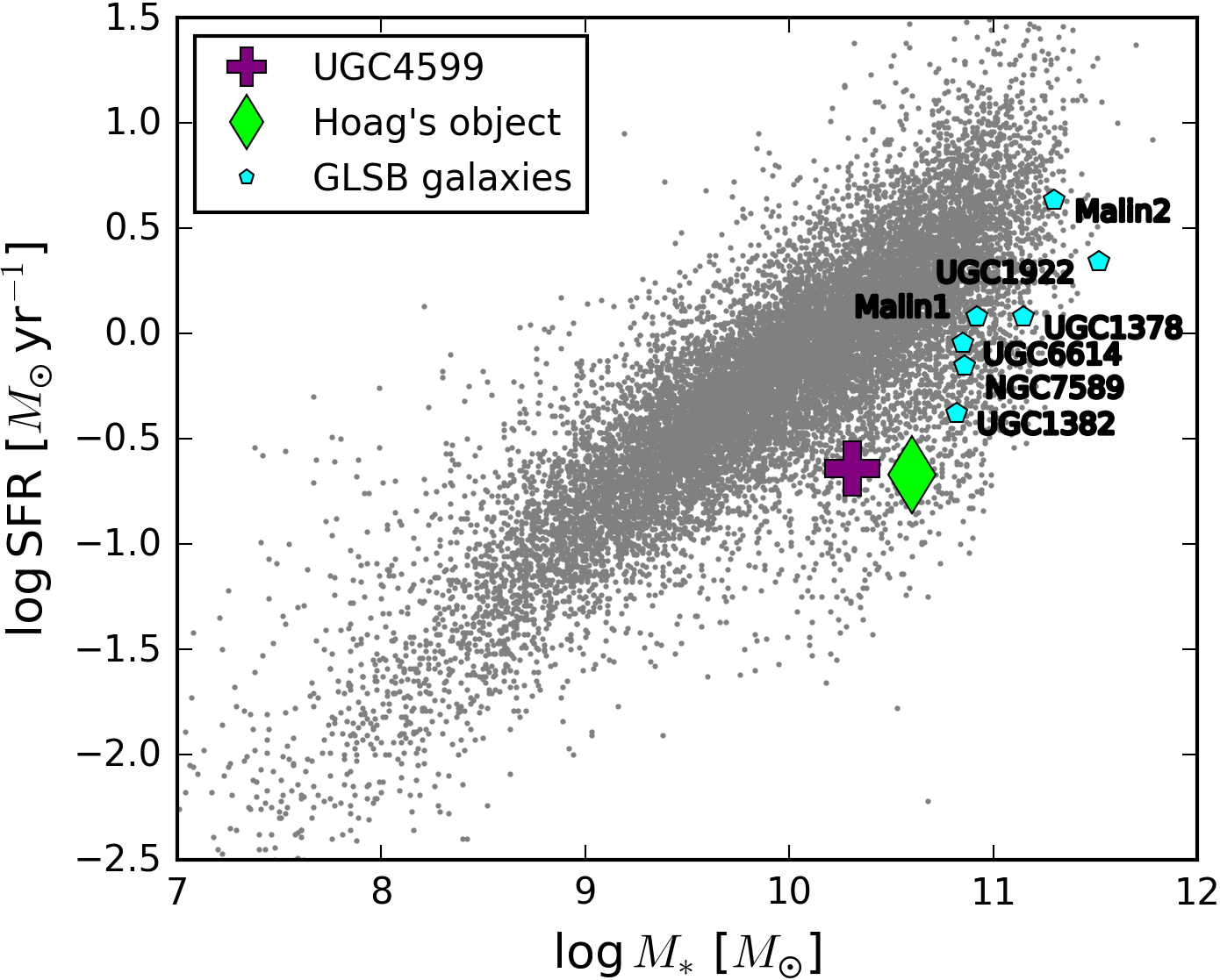}
	\includegraphics[width=0.495\textwidth]{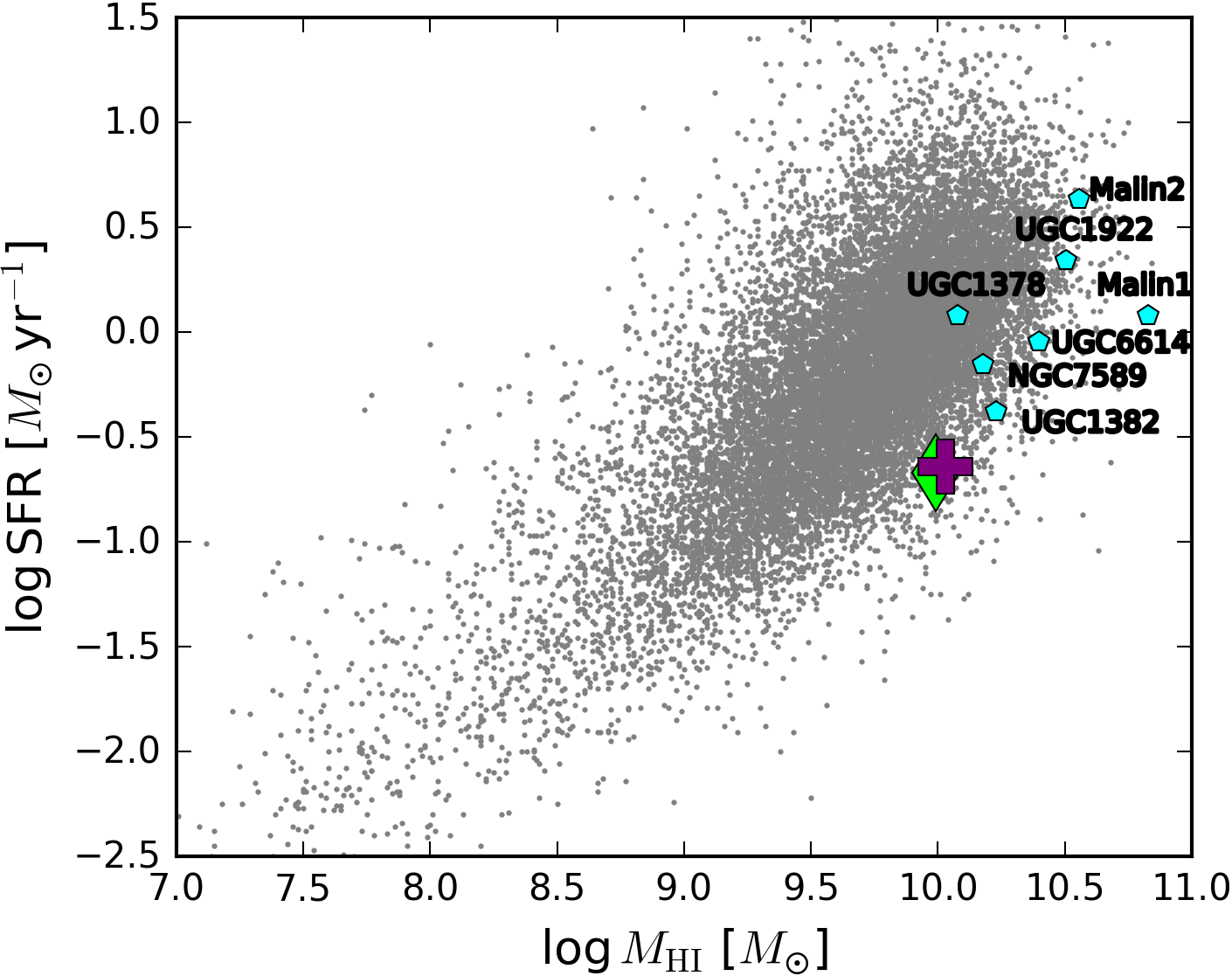}
		\caption{SFR as a function of stellar mass ({\it left}) and as a function of mass of neutral hydrogen ({\it right}). Grey dots depict galaxies from the ALFALFA-SDSS Galaxy Catalog \citep{2020AJ....160..271D} where the SFR was estimated using GALEX NUV photometry and the stellar mass was calculated using the method proposed by \citet{2015ApJ...802...18M}.}
	\label{M_SFR}
\end{figure*}

In Fig.~\ref{M_SFR}, we consider two correlations between the SFR and the stellar mass ({\it left} plot) and the SFR and the neutral gas mass ({\it right} plot). As can be seen, UGC\,4599, Hoag's object, and the GLSB galaxies are obvious outliers from these relations, implying that these galaxies have a significantly lower SFR than regular galaxies with the same stellar or gas mass.

\subsection{Formation scenarios of UGC\,4599}
\label{sec:formation_scenarios}
The formation of this intriguing galaxy may be explained by several scenarios, discussed in detail in \citet{2011MNRAS.413.2621F} and \citet{2023A&A...669L..10S}. Here, we only list most plausible of them, in line with our analysis. The collisional ring scenario, major merging, minor accretion events, and the external accretion of material from a donor galaxy have been largely discarded because they are not supported by observations.

Firstly, the formation of the star-forming ring in UGC\,4599 may be related to secular evolution through bar-related resonances \citep{1996FCPh...17...95B}. Our $u$-band image in Fig.~\ref{fig:uvimg} showcases at least one inner spiral arm with a larger pitch angle than in the outer structure, winding up toward the nucleus. Also, in contrast to the conclusion in \citet{2011MNRAS.413.2621F} that the bulge in UGC\,4599 is de Vaucouleurs, we find that it is exponential (or close to it), so it should be classified as a pseudobulge (\citealt{2008AJ....136..773F,2009MNRAS.393.1531G} and references therein). Pseudobulges hold clues to the formation of bars and nuclear structures (nuclear bars, nuclear rings, nuclear spirals, and other nuclear substructures) through secular processes in the disks and bulges \citep{2004ARA&A..42..603K}. Therefore, the morphology of UGC\,4599 is probably governed by secular evolution. Perhaps, high-resolution observations will reveal nuclear substructures which are smeared out in our ground-based images.

Secondly, the existence of the star-forming ring, the extremely faint XUV-disk, and even more extended and similar (in mass) hydrogen disk suggest an external origin of the gas through the cold accretion from the intergalactic medium. This is a likely scenario because most galaxies in the group where UGC\,4599 resides are gas-rich and blue, with star-forming or starburst spectra (e.g. CGCG~061-011 or MCG+02-22-006). Also, the time for accumulating the estimated amount of gas in UGC\,4599 (the accretion rate depends on the halo mass) is consistent with the estimate on the age of UGC\,4599 (see discussion in \citealt{2011MNRAS.413.2621F}). Most importantly, this scenario is supported by recent results 
from \citet{2023A&A...669L..10S} who used long-slit spectroscopy and other data to explore the stellar kinematics in the central body of UGC\,4599 and different strong emission-line flux ratios in the ring. Surprisingly, the gas metallicity in the ring appeared to be unusually low as compared to the majority of the outer star-forming rings in S0 galaxies with nearly solar metallicities \citep{2019ApJS..244....6S,2020A&A...634A.102P}. Therefore, based on our deep photometric analysis and recent results by \citet{2023A&A...669L..10S}, we conclude that the ring and the XUV disk of UGC~4599 are likely the result of gas accretion from a cosmological filament. 
Interestingly, Hoag's Object is believed to be formed through the same mechanism \citep{2011MNRAS.418.1834F,2013MNRAS.435..475B}, but instead of a giant LSB disk with an extended spiral structure only a star-forming ring around the elliptical galaxy was produced. In contrast to UGC\,4599, Hoag's Object generally follows the main galaxy scaling relations.

\section{Conclusions}
\label{sec:conclusions}	

This research more closely examines the structure of the GLSB galaxy UGC\,4599 consisting of a reddish core, blue star-forming ring, and an extremely faint outer disk with embedded two-armed spiral structure. We have used several data sources from the UV to NIR, including our own deep observations in the $u$ band on the 4.3m Lowell Discovery Telescope and in the wide Luminance filter using the 0.7m Jeanne Rich telescope in the framework of the {\sl HERON} survey reaching a surface brightness limit of 31~mag\,arcsec$^{-2}$. We have also analyzed a VLA H{\sc i} map of UGC\,4599 taken from the literature.

We report that the two-armed structure manifests itself in both UV and deep optical observations and extends out to 45~kpc. This structure is also well separated in the H{\sc i} disk. As such, UGC\,4599 demonstrates many of the characteristics of XUV disks \citep{2007ApJS..173..538T,2010ApJ...714L.290S}, which can be detected not only in UV observations, but also in deep optical integrations. 

The mean pitch angle for the two-armed outer spiral structure in UGC\,4599 is estimated to be $P=6.4^{\circ} \pm1.5^{\circ}$. The spiral structure of this galaxy resembles the one in other GLSB galaxies: it exhibits relatively narrow spiral arms, has a small pitch angle, and winds multiple times about the galaxy center. The spiral arms are extremely faint ($\sim25-27$~mag\,arcsec$^{-2}$ in the {\sl HERON} image) and become apparent only in deep UV and optical images. Based on our deep $u$-band image, we trace the inner spiral structure within the galaxy core, in contrast to \citet{2011MNRAS.413.2621F} who claim that the ring is completely detached from the central component. The inner spiral arm has a larger pitch angle than the overall outer spiral structure and may point to the existence of a bar in the near past. The parameters of the stellar spiral structure and the one traced by the neutral hydrogen are, on average, well consistent but show some differences at smaller scales. 

We undertook a detailed investigation of the structure of UGC\,4599 to characterize how this galaxy compares to ordinary spiral galaxies and other GLSB galaxies. We fitted a single exponential disk to azimuthally averaged profiles obtained in different wavelengths, from the UV to 3.6~$\mu$m. The disk scale length decreases with wavelength and differs by a factor of three for the young (GALEX FUV) and old ({\it Spitzer} IRAC 3.6~$\mu$m) stellar populations. The gaseous disk has the largest extent with a disk scale length of 25~kpc. Our accurate 2D decomposition of the {\sl HERON} image reveals an unresolved nuclear component, an exponential pseudobulge with a low effective surface brightness ($\mu_\mathrm{e,b}(r)=22.07$~mag\,arcsec$^{-2}$) but a large effective radius ($r_\mathrm{e,b}=1.67$~kpc). The outer stellar disk has an extremely low central surface brightness ($\mu_\mathrm{0,d}(r)=25.5$~mag\,arcsec$^{-2}$) and a very large scale length  ($h=15.0$~kpc), comparable to the disk parameters in GLSB galaxies.

The comparison of UGC\,4599 with regular early- and late-type galaxies, as well as with other GLSB galaxies and Hoag's Object (Fig.~\ref{fig:cors}), on the Kormendy relation, bulge size--luminosity relation, the fundamental plane for bulges and elliptical galaxies, and the disk surface brightness--scale length relation shows that on each of these relations UGC\,4599 significantly deviates from other galaxies, including the selected GLSB galaxies. Hoag's Object, on the contrary, follows the general trends for the elliptical galaxies. 

Finally, we measured the SFR of the entire galaxy and its XUV disk. A majority of the SFR occurs in the outer disk and embedded spiral arms (77\%) rather than in the pseudobulge and the ring (23\%), whereas the total SFR is $0.229 M_\odot\,\mathrm{yr}^{-1}$ which is typical of average-sized LSB galaxies but slightly lower than in GLSB galaxies.  

Based on our analysis and previous studies of UGC\,4599, we conclude that cold accretion of gas through a cosmological filament is the most favoured mechanism for the formation of the observed extended, LSB outer disk with embedded two-armed spiral structure. The star-forming ring within the inner disk may have formed thanks to the same source of gas via secular evolution.


\section*{Acknowledgements}
We thank Olga Sil'chenko for her comments, which allowed us to improve the quality of the publication.

Alexander Marchuk and Aleksandr Mosenkov acknowledge financial support from the Russian Science Foundation (grant no. 22-22-00483).

These results made use of the 4.3-meter Lowell Discovery Telescope (LDT), formerly the Discovery Channel Telescope (DCT). Lowell is a private, non-profit institution dedicated to astrophysical research and public appreciation of astronomy and operates the LDT in partnership with Boston University, the University of Maryland, the University of Toledo, Northern Arizona University and Yale University. The Large Monolithic Imager was built by Lowell Observatory using funds provided by the National Science Foundation (AST-1005313). This research was funded in part by NASA through the Arkansas Space Grant Consortium. The authors acknowledge logistical support from the Polaris Observatory Association in the operations of the Jeanne Rich 0.7m telescope. We would like to thank Jayce Dowell and Liese van Zee for their input and directing us to Neutral Hydrogen data.

This research has made use of the NASA/IPAC Infrared Science Archive (IRSA; \url{http://irsa.ipac.caltech.edu/frontpage/}), and the NASA/IPAC Extragalactic Database (NED; \url{https://ned.ipac.caltech.edu/}), both of which are operated by the Jet Propulsion Laboratory, California Institute of Technology, under contract with the National Aeronautics and Space Administration.  This research has made use of the HyperLEDA database (\url{http://leda.univ-lyon1.fr/}; \citealp{2014A&A...570A..13M}). 
This work is based in part on observations made with the {\it Spitzer} Space Telescope, which is operated by the Jet Propulsion Laboratory, California Institute of Technology under a contract with NASA. 

This project used data obtained with the Dark Energy Camera (DECam), which was constructed by the Dark Energy Survey (DES) collaboration. Funding for the DES Projects has been provided by the U.S. Department of Energy, the U.S. National Science Foundation, the Ministry of Science and Education of Spain, the Science and Technology Facilities Council of the United Kingdom, the Higher Education Funding Council for England, the National Center for Supercomputing Applications at the University of Illinois at Urbana-Champaign, the Kavli Institute of Cosmological Physics at the University of Chicago, Center for Cosmology and Astro-Particle Physics at the Ohio State University, the Mitchell Institute for Fundamental Physics and Astronomy at Texas A\&M University, Financiadora de Estudos e Projetos, Fundacao Carlos Chagas Filho de Amparo, Financiadora de Estudos e Projetos, Fundacao Carlos Chagas Filho de Amparo a Pesquisa do Estado do Rio de Janeiro, Conselho Nacional de Desenvolvimento Cientifico e Tecnologico and the Ministerio da Ciencia, Tecnologia e Inovacao, the Deutsche Forschungsgemeinschaft and the Collaborating Institutions in the Dark Energy Survey. The Collaborating Institutions are Argonne National Laboratory, the University of California at Santa Cruz, the University of Cambridge, Centro de Investigaciones Energeticas, Medioambientales y Tecnologicas-Madrid, the University of Chicago, University College London, the DES-Brazil Consortium, the University of Edinburgh, the Eidgenossische Technische Hochschule (ETH) Zurich, Fermi National Accelerator Laboratory, the University of Illinois at Urbana-Champaign, the Institut de Ciencies de l'Espai (IEEC/CSIC), the Institut de Fisica d'Altes Energies, Lawrence Berkeley National Laboratory, the Ludwig-Maximilians Universitat Munchen and the associated Excellence Cluster Universe, the University of Michigan, the National Optical Astronomy Observatory, the University of Nottingham, the Ohio State University, the University of Pennsylvania, the University of Portsmouth, SLAC National Accelerator Laboratory, Stanford University, the University of Sussex, and Texas A\&M University.

The Legacy Surveys consist of three individual and complementary projects: the Dark Energy Camera Legacy Survey (DECaLS; NOAO Proposal ID \# 2014B-0404; PIs: David Schlegel and Arjun Dey), the Beijing-Arizona Sky Survey (BASS; NOAO Proposal ID \# 2015A-0801; PIs: Zhou Xu and Xiaohui Fan), and the Mayall z-band Legacy Survey (MzLS; NOAO Proposal ID \# 2016A-0453; PI: Arjun Dey). DECaLS, BASS and MzLS together include data obtained, respectively, at the Blanco telescope, Cerro Tololo Inter-American Observatory, National Optical Astronomy Observatory (NOAO); the Bok telescope, Steward Observatory, University of Arizona; and the Mayall telescope, Kitt Peak National Observatory, NOAO. The Legacy Surveys project is honored to be permitted to conduct astronomical research on Iolkam Du'ag (Kitt Peak), a mountain with particular significance to the Tohono O'odham Nation.

NOAO is operated by the Association of Universities for Research in Astronomy (AURA) under a cooperative agreement with the National Science Foundation.

BASS is a key project of the Telescope Access Program (TAP), which has been funded by the National Astronomical Observatories of China, the Chinese Academy of Sciences (the Strategic Priority Research Program "The Emergence of Cosmological Structures" Grant \# XDB09000000), and the Special Fund for Astronomy from the Ministry of Finance. The BASS is also supported by the External Cooperation Program of Chinese Academy of Sciences (Grant \# 114A11KYSB20160057), and Chinese National Natural Science Foundation (Grant \# 11433005).

The Legacy Survey team makes use of data products from the Near-Earth Object Wide-field Infrared Survey Explorer (NEOWISE), which is a project of the Jet Propulsion Laboratory/California Institute of Technology. NEOWISE is funded by the National Aeronautics and Space Administration.

The Legacy Surveys imaging of the DESI footprint is supported by the Director, Office of Science, Office of High Energy Physics of the U.S. Department of Energy under Contract No. DE-AC02-05CH1123, by the National Energy Research Scientific Computing Center, a DOE Office of Science User Facility under the same contract; and by the U.S. National Science Foundation, Division of Astronomical Sciences under Contract No. AST-0950945 to NOAO.

This research has made use of GALEX data obtained from the Mikulski Archive for Space Telescopes (MAST); support for MAST for non-HST data is provided by the NASA Office of Space Science via grant NNX09AF08G and by other grants and contracts (MAST is maintained by STScI, which is operated by the Association of Universities for Research in Astronomy, Inc., under NASA contract NAS5-26555).

This research has made use of Karl G. Jansky Very Large Array (VLA) data which were kindly provided by Jayce Dowell. The National Radio Astronomy Observatory is a facility of the National Science Foundation operated under cooperative agreement by Associated Universities, Inc.

This research made use of the ``K-corrections calculator'' service available at http://kcor.sai.msu.ru/
%

\section*{Data Availability}

The data underlying this article will be shared on reasonable request to the corresponding author.



\bibliographystyle{mnras}
\bibliography{art} 

\appendix
\section{2D photometric decomposition of Hoag's Object}
\label{sec:hoag}

We exploited the DESI Legacy Imaging Surveys \citep{2019AJ....157..168D} for fitting the structure of Hoag's Object in the $r$ band. A cut-out with the galaxy was retrieved from the Legacy viewer, along with a coadd PSF FITS file. We corrected for the flat background and masked off all other background and foreground sources. To fully take into account the scattered light from the PSF, we created an extended PSF image, with the core described by the retrieved coadd PSF image (within $r<8$~arcsec) and the outer part of the profile (the wings, $r\geq8$~arcsec) using a fixed power law\footnote{See \url{https://www.legacysurvey.org/dr9/psf/} for details.} $r^{-2}$ out to $r=100$~arcsec. 

\begin{figure}
\includegraphics[width=0.46\textwidth]{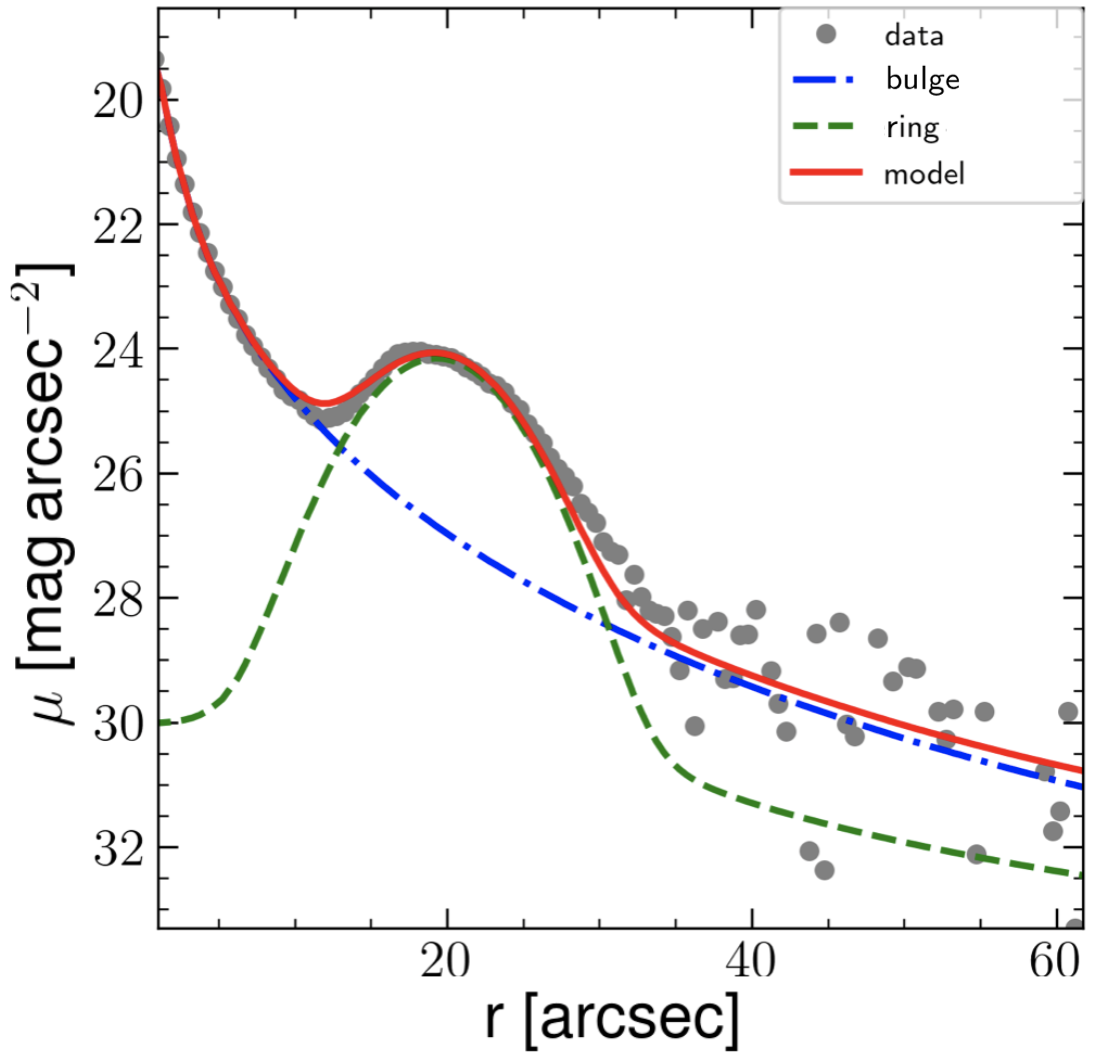}
\caption{{\tt{IMFIT}} fit of Hoag's Object using the $r$-band image taken from the DESI Legacy Imaging Surveys. The components include a central bulge and a star forming ring.}
\label{azim_hoag}
\centering
\end{figure}

\begin{figure*}
\hspace{8mm}
\includegraphics[width=\textwidth]{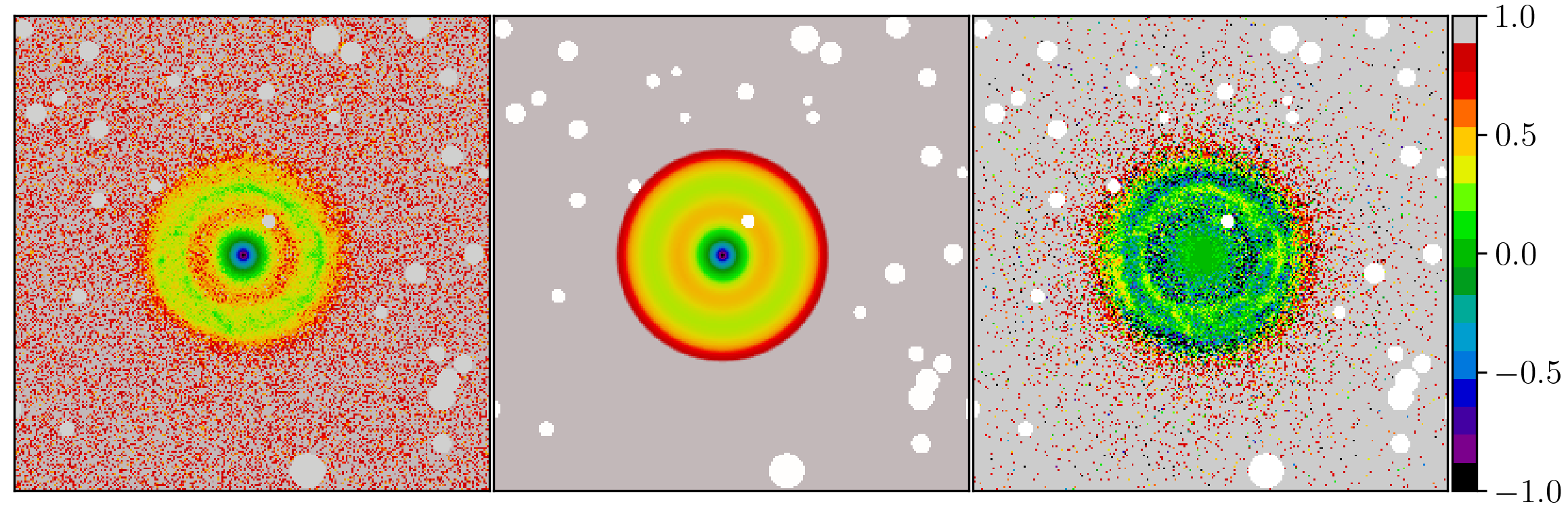}
\caption{{\it Left}: Image of Hoag's Object taken from the DESI Legacy Imaging Surveys with masked foreground and background objects. {\it Middle}: {\tt{IMFIT}} model fit of Hoag's Object.
{\it Right}: Relative residual image of (Observed Image-Model)/Observed Image.}
\label{plot2d_hoag}
\centering
\end{figure*}

The decomposition was carried out into S\'ersic and Gaussian ring components using the {\tt{IMFIT}} code. The results of the fitting are listed in Table~\ref{tab:hoag}. The Figs.~\ref{azim_hoag} and \ref{plot2d_hoag} demonstrate a 1D azimuthally averaged profile and 2D image/model/residual images, respectively. As one can see, the profile is extending down to 31~mag\,arcsec$^{-2}$ and is described by the model fairly well. We see no signs of a stellar disk or a halo. The Gaussian ring adequately recovers the bump in the 1D profile, which corresponds to the tightly wound spiral arms in the galaxy image. The parameters of the S\'ersic profile are consistent with \citet{2011MNRAS.418.1834F} who fitted a $B$-band surface brightness profile with a two-component `S\'ersic + exponential disk' model: $n=3.9\pm0.2$ (versus our $4.1$), $r_\mathrm{e,b}=2.8\pm0.1$~arcsec (versus our $2.5$~arcsec), and $\mu_{\mathrm{e,b}}(B)=22.6\pm0.5$~mag\,arcsec$^{-2}$ (versus our $21.23$~mag\,arcsec$^{-2}$ in the $r$ band). The $r$-band deep image and, most importantly, the residual image do not reveal any (LSB) tidal features suggesting that the galaxy is in a quasi-equilibrium state.

\begin{table}
\caption{Results of the {\tt{IMFIT}} modeling for Hoag's Object. Under each component's name we provide the corresponding {\tt{IMFIT}} function used.}
\label{tab:hoag}
\centering
    \begin{tabular}{lccc}
    \hline
    \hline\\[-1ex]    
    Component & Parameter &  Value & Units \\[0.5ex]
    \hline\\[-0.5ex]
    1. Bulge:        & $n$  &  $4.1$  &  \\[+0.5ex]
    ({\it S\'ersic}) & $\mu_{\mathrm{e,b}}$ & $21.23$ & mag\,arcsec$^{-2}$  \\[+0.5ex] 
                     & $r_\mathrm{e,b}$ & $2.22$ & kpc \\[+0.5ex]  
                     & $M_\mathrm{b}$ &  $-20.61$ & mag  \\[+0.5ex]
                     & $f_\mathrm{b}$     & $0.599$     &          \\[+0.5ex]
    2. Ring:         & $\mu_\mathrm{r}$  &  $25.45$  & mag\,arcsec$^{-2}$ \\[+0.5ex]
 ({\it GaussianRing})& $R_\mathrm{r}$ & $26.89$ & kpc \\[+0.5ex]  
                     & $\sigma_\mathrm{r}$ & $3.32$ & kpc \\[+0.5ex]
                     & $M_\mathrm{r}$ &  $-20.17$ & mag  \\[+0.5ex]
                     & $f_\mathrm{r}$     & $0.401$     &          \\[+0.5ex]
    \hline\\[-0.5ex]
    \end{tabular}
     \parbox[t]{85mm}{
    \textbf{Notes:} $n$ is the S\'ersic index, $\mu_{\mathrm{e,b}}$ is the effective radius of the S\'ersic component at the effective radius $r_\mathrm{e,b}$; $\mu_\mathrm{r}$ is the maximum surface brightness of the elliptical Gaussian ring centered at $R_\mathrm{r}$ with the $\sigma_\mathrm{r}$ parameter controlling the width of the ring. $M$ and $f$ denote the absolute magnitude and the fraction of the component's luminosity, respectively. The surface brightnesses and absolute magnitudes have been corrected for Galactic extinction using \citet{2011ApJ...737..103S} and K-correction using the K-correction calculator \citep{2010MNRAS.405.1409C,2012MNRAS.419.1727C}.}
\end{table}

\bsp	
\label{lastpage}
\end{document}